\documentclass[twocolumn]{aastex631}

\graphicspath{{figure/}}
\usepackage{macros}
\usepackage{hyperref}

\defcitealias{Wetzel2016}{W16}
\defcitealias{Chan2018}{C18}
\defcitealias{ElBadry2018a}{E18}
\defcitealias{Hopkins2018b}{H18}
\defcitealias{Wheeler2019}{W19}
\defcitealias{GarrisonKimmel2017}{G17}
\defcitealias{GarrisonKimmel2019a}{G19a}
\defcitealias{GarrisonKimmel2019b}{G19b}
\defcitealias{Samuel2020}{S20}
\defcitealias{AnglesAlcazar2017b}{AA17}
\defcitealias{Ma2018a}{M18}
\defcitealias{Ma2019}{M19}
\defcitealias{Ma2020b}{M20}

\newcommand{\ucdavis}{Department of Physics \& Astronomy, University of California, Davis, CA 95616, USA}
\newcommand{\upenn}{Department of Physics \& Astronomy, University of Pennsylvania, 209 S 33rd St, Philadelphia, PA 19104, USA}
\newcommand{\cca}{Center for Computational Astrophysics, Flatiron Institute, 162 5th Ave, New York, NY 10010, USA}
\newcommand{\uconn}{Department of Physics, University of Connecticut, 196 Auditorium Road, U-3046, Storrs, CT 06269-3046, USA}
\newcommand{\uarizona}{Steward Observatory, University of Arizona, 933 N Cherry Ave, Tucson, AZ 85719, USA}
\newcommand{\zurich}{Institute for Computational Science, University of Zurich, Winterthurerstrasse 190, Zurich CH-8057, Switzerland}

\newcommand{\uchicago}{Department of Astronomy \& Astrophysics, University of Chicago, Chicago, IL 60637, USA}
\newcommand{\cfa}{Center for Astrophysics, Harvard \& Smithsonian, 60 Garden Street, Cambridge, MA 02138, USA}
\newcommand{\hsf}{Harvard Society of Fellows, 78 Mount Auburn Street, Cambridge, MA 02138, USA}
\newcommand{\calpoly}{Department of Physics \& Astronomy, California State Polytechnic University, Pomona, CA 91768, USA}
\newcommand{\caltech}{
TAPIR, Mailcode 350-17, California Institute of Technology, Pasadena, CA 91125, USA}
\newcommand{\ucmerced}{Department of Physics, University of California, Merced, 5200 N Lake Road, Merced, CA 95343, USA}
\newcommand{\utaustin}{Department of Astronomy, The University of Texas at Austin, 2515 Speedway, Stop C1400, Austin, TX 78712-1205, USA}
\newcommand{\columbia}{Department of Astronomy, Columbia University, New York, NY 10027, USA}
\newcommand{\ucirvine}{Department of Physics \& Astronomy, 4129 Reines Hall, University of California, Irvine, CA 92697, USA}
\newcommand{\northwestern}{Department of Physics \& Astronomy and CIERA, Northwestern University, 1800 Sherman Ave, Evanston, IL 60201, USA}
\newcommand{\ucsd}{Center for Astrophysics \& Space Sciences, University of California, San Diego, 9500 Gilman Drive, La Jolla, CA 92093, USA}
\newcommand{\princeton}{Department of Astrophysical Sciences, Princeton University, Peyton Hall, Princeton, NJ 08544, USA}

\begin{document}

\title{Public data release of the FIRE-2 cosmological zoom-in simulations of galaxy formation}
\shorttitle{FIRE-2 public data release}
\shortauthors{Wetzel et al.}

\correspondingauthor{Andrew Wetzel}
\email{awetzel@ucdavis.edu}

\author[0000-0003-0603-8942]{Andrew Wetzel}
\affiliation{\ucdavis}

\author[0000-0003-4073-3236]{Christopher C. Hayward}
\affiliation{\cca}

\author[0000-0003-3939-3297]{Robyn E. Sanderson}
\affiliation{\cca}
\affiliation{\upenn}

\author{Xiangcheng Ma}
\affiliation{\uarizona}

\author[0000-0001-5769-4945]{Daniel Angl{\'e}s-Alc{\'a}zar}
\affiliation{\uconn}
\affiliation{\cca}

\author[0000-0002-1109-1919]{Robert Feldmann}
\affiliation{\zurich}

\author[0000-0003-2544-054X]{T.K Chan}
\affiliation{\uchicago}

\author{Kareem El-Badry}
\affiliation{\cfa}
\affiliation{\hsf}

\author{Coral Wheeler}
\affiliation{\calpoly}

\author{Shea Garrison-Kimmel}
\affiliation{\caltech}

\author[0000-0002-3641-4366]{Farnik Nikakhtar}
\affiliation{\upenn}

\author[0000-0001-5214-8822]{Nondh Panithanpaisal}
\affil{\upenn}

\author[0000-0002-8354-7356]{Arpit Arora}
\affiliation{\upenn}

\author[0000-0002-6145-3674]{Alexander B. Gurvich}
\affiliation{\northwestern}

\author[0000-0002-8429-4100]{Jenna Samuel}
\affiliation{\utaustin}
\affiliation{\ucdavis}

\author{Omid Sameie}
\affiliation{\utaustin}

\author{Viraj Pandya}
\affiliation{\columbia}

\author[0000-0001-7326-1736]{Zachary Hafen}
\affiliation{\ucirvine}

\author[0000-0002-3817-8133]{Cameron Hummels}
\affiliation{\caltech}

\author[0000-0003-3217-5967]{Sarah Loebman}
\affiliation{\ucmerced}

\author[0000-0002-9604-343X]{Michael Boylan-Kolchin}
\affiliation{\utaustin}

\author[0000-0003-4298-5082]{James S. Bullock}
\affiliation{\ucirvine}

\author[0000-0002-4900-6628]{Claude-Andr{\'e} Faucher-Gigu{\`e}re}
\affiliation{\northwestern}

\author{Du\v{s}an Kere\v{s}}
\affiliation{\ucsd}

\author{Eliot Quataert}
\affiliation{\princeton}

\author[0000-0003-3729-1684]{Philip F. Hopkins}
\affiliation{\caltech}

\begin{abstract}
We describe a public data release of the FIRE-2 cosmological zoom-in simulations of galaxy formation, available at \href{http://flathub.flatironinstitute.org/fire}{flathub.flatironinstitute.org/fire}, from the Feedback In Realistic Environments (FIRE) project.
FIRE-2 simulations achieve parsec-scale resolution to explicitly model the multiphase interstellar medium while implementing direct models for stellar evolution and feedback, including stellar winds, core-collapse and Ia supernovae, radiation pressure, photoionization, and photoelectric heating.
We release complete snapshots from 3 suites of simulations.
The first comprises 20 simulations that zoom in on 14 Milky Way-mass galaxies, 5 SMC/LMC-mass galaxies, and 4 lower-mass galaxies including 1 ultrafaint; we release 39 snapshots across $z\!=\!0\!-\!10$.
The second comprises 4 massive galaxies, with 19 snapshots across $z\!=\!1\!-\!10$.
Finally, a high-redshift suite comprises 22 simulations, with 11 snapshots across $z\!=\!5\!-\!10$.
Each simulation also includes dozens of resolved lower-mass (satellite) galaxies in its zoom-in region.
Snapshots include all stored properties for all dark matter, gas, and star particles, including 11 elemental abundances for stars and gas, and formation times (ages) of star particles.
We also release accompanying (sub)halo catalogs, which include galaxy properties and member star particles.
For the simulations to $z\!=\!0$, including all Milky Way-mass galaxies, we release the formation coordinates and an ``ex-situ" flag for all star particles, pointers to track particles across snapshots, catalogs of stellar streams, and multipole basis expansions for the halo mass distributions.
We describe publicly available python packages for reading and analyzing these simulations.
\end{abstract}

~

\section{Introduction}
\label{sec:intro}

Cosmological simulations that model the physics of both dark matter and baryons (gas plus stars) are powerful tools to understand the formation of dark-matter (sub)halos and their galaxies \citep[for recent reviews, see][]{Vogelsberger2020, Sales2022}.
Cosmological simulations that zoom in on a region around a galaxy \citep{KatzWhite1993, HahnAbel2011, Onorbe2014} provide the highest resolution and can resolve individual star-forming regions and stellar populations; in some cases they now achieve mass resolution comparable to individual massive stars.

Over the last decade, several groups have generated various cosmological zoom-in baryonic simulations that span the range of galaxy masses and redshifts, from the lowest-mass galaxies \citep[for example][]{Wang2015, Xu2016, Jeon2017, Ceverino2017, Fitts2017, Ma2018a, Revaz2018, Wheeler2019, Rey2019, Munshi2021}, to \ac{MW}-mass galaxies \citep[for example][]{Guedes2011, Sawala2016b, Wetzel2016, Grand2017, Buck2019, GarrisonKimmel2019a, Peeples2019, Libeskind2020, Agertz2021, Applebaum2021, Font2021}, to the most massive galaxies and galaxy clusters \citep[for example][]{Nelson2014, Wu2015, Feldmann2016, AnglesAlcazar2017b, Bahe2017, Barnes2017b, Tremmel2019}.

One collaboration developing cosmological baryonic simulations is the Feedback In Realistic Environments (FIRE) project\footnote{
\href{https://fire.northwestern.edu}{fire.northwestern.edu}
}
\citep[introduced in][]{Hopkins2014a}.
The FIRE project seeks to develop cosmological simulations of galaxy formation that resolve the multiphase \ac{ISM}, while implementing all of the major channels for stellar feedback from stellar evolution models as directly as possible, within a cosmological context.
By achieving parsec-scale resolution and anchoring the feedback prescriptions directly to stellar population models, FIRE aims to improve the predictive power of cosmological simulations of galaxy formation.

\citet{Hopkins2014a} introduced the first-generation FIRE-1 physics model and a suite of FIRE-1 simulations (originally named simply the FIRE simulations), while \citet{Hopkins2018b} introduced the second-generation FIRE-2 physics model and an initial suite of FIRE-2 simulations.
The key improvements in FIRE-2 over FIRE-1 were primarily numerical: (1) pushing to higher resolution, (2) switching from \ac{SPH} to the more accurate \ac{MFM} Godunov method for hydrodynamics, (3) using a more accurate, geometrically-aware method for coupling mechanical feedback from stars to surrounding gas, (4) increasing the density threshold for star formation from $n\!>\!50 \cci$ to $n\!>\!1000 \cci$, and (5) adding an explicit model for subgrid mixing/diffusion of metals in gas via turbulence.
More recently, \citet{Hopkins2023} introduced the FIRE-3 model, whose key improvements focus on the underlying models for stellar evolution and gas cooling at low temperatures.

To date, FIRE-2 simulations have been used in over 100 publications that explore numerous facets of galaxy formation.
As some examples, we used the FIRE-2 simulations that include the base set of physics (which we publicly release and describe below) to examine:
the formation of low-mass galaxies \citep[for example][]{Fitts2017, Chan2018, Wheeler2019} including during the epoch of reionization \citep{Ma2018a} and as satellites of \ac{MW}-mass galaxies \citep[for example][]{Wetzel2016, GarrisonKimmel2019a};
the formation of massive galaxies and black hole growth \citep[for example][]{AnglesAlcazar2017b};
the \ac{ISM} \citep[for example][]{ElBadry2017, Gurvich2020, Orr2020}, including galactic winds \citep{Pandya2021}, giant molecular clouds \citep{Benincasa2020, Guszejnov2020}, and star clusters \citep{Ma2020a};
the circum-galactic medium \citep[for example][]{Hafen2019, Hafen2020, Stern2021};
star formation \citep[for example][]{FloresVelazquez2021};
disk formation \citep[for example][]{GarrisonKimmel2018, Santistevan2020, Yu2021};
elemental abundance distributions in stars and gas \citep[for example][]{Escala2018, Bellardini2021};
stellar halos \citep[for example][]{Bonaca2017, Sanderson2018};
dark matter within galaxies \citep[for example][]{Necib2019, Lazar2020} and in surrounding subhalos \citep[for example][]{GarrisonKimmel2017};
and models for binary black hole populations \citep{Lamberts2018}.

This article describes the first full public data release (DR1) of the FIRE-2 cosmological zoom-in simulations, available at \href{http://flathub.flatironinstitute.org/fire}{flathub.flatironinstitute.org/fire}.
This release includes 3 suites of simulations: a \textit{Core} suite to $z\!=\!0$, a \textit{Massive Halo} suite to $z\!=\!1$ \citep{AnglesAlcazar2017b}, and a \textit{High Redshift} suite to $z\!=\!5$ \citep{Ma2018a}.
This DR1 extends our initial data release (DR0) of a subset of FIRE-2 simulations, which contained complete snapshots of 3 simulations of \ac{MW}-mass galaxies at $z\!=\!0$ (m12f, m12i, and m12m, all included here), accompanied by 9 \textsc{Ananke} synthetic Gaia DR2-like surveys that we created from these simulations \citep{Sanderson2020}, which are hosted via \textit{yt Hub} at \href{https://ananke.hub.yt}{ananke.hub.yt}.

FIRE-2 DR1 represents the first public data release of a suite of cosmological zoom-in baryonic simulations across cosmic time.
It adds to the existing set of public data releases of larger-volume, uniform-resolution cosmological baryonic simulations to $z\!=\!0$, such as Illustris \citep{Nelson2015b} and Illustris TNG \citep{Nelson2019}, EAGLE \citep{McAlpine2016}, Simba \citep{Dave2019}, and CAMELS \citep{VillaescusaNavarro2022}.
Thus, a user can compare and/or combine FIRE-2 with these larger-volume, but lower-resolution, cosmological simulations.

\section{FIRE-2: methods and caveats}
\label{sec:method}

\subsection{FIRE-2 model}
\label{sec:model}

We generated all simulations using \textsc{Gizmo}\footnote{
\href{http://www.tapir.caltech.edu/~phopkins/Site/GIZMO.html}{www.tapir.caltech.edu/$\sim$phopkins/Site/GIZMO.html}
} \citep{Hopkins2015}, a multimethod gravity plus (magneto)hydrodynamics code.
We used the mesh-free finite-mass (MFM) mode for hydrodynamics, a quasi-Lagrangian Godunov method that provides adaptive spatial resolution while maintaining exact conservation of mass, energy, and momentum, excellent angular momentum conservation, and accurate shock capturing.
Thus, the method provides advantages of both smoothed-particle hydrodynamics (SPH) and Eulerian adaptive mesh refinement (AMR) methods.
\textsc{Gizmo} solves gravity using an improved version of the Tree-PM solver from GADGET-3 \citep{Springel2005e}, using fully adaptive and conservative gravitational force softening for gas cells that matches their hydrodynamic resolution.

All of these simulations use the same FIRE-2 physics model \citep{Hopkins2018b}, with minor exceptions that we describe below.
Briefly, FIRE-2 incorporates radiative cooling and heating across $10 - 10^{10}$ K, including free-free, photoionization and recombination, Compton, photoelectric and dust collisional, cosmic ray, molecular, metal-line, and fine-structure processes, self-consistently tracking 11 elements (H, He, C, N, O, Ne, Mg, Si, S, Ca, Fe).
This includes photoionization and heating from a redshift-dependent, spatially uniform ultraviolet background \citep{FaucherGiguere2009}, which reionizes the universe at $z \approx 10$.\footnote{
The exact model used is available at \\ \href{https://galaxies.northwestern.edu/uvb-fg09}{galaxies.northwestern.edu/uvb-fg09}}
The modeling of ionization also includes approximations for self-shielding of dense gas and radiation from local sources based on the LEBRON scheme \citep{Hopkins2020a}.

Star formation occurs in \textit{self-gravitating} gas \citep[following][]{Hopkins2013c} that also is molecular and self-shielding \citep[following][]{KrumholzGnedin2011}, Jeans unstable, and exceeds a minimum density threshold, $n_{\rm SF}\!>\!1000 \cci$.
FIRE-2 follows several stellar-feedback mechanisms, including:
(1) local and long-range momentum flux from radiation pressure in the ultraviolet and optical (single-scattering), as well as reradiated light in the infrared; (2) energy, momentum, mass and metal injection from core-collapse + Ia supernovae and stellar mass loss (dominated by O, B, and AGB stars); and (3) photoionization and photoelectric heating.
FIRE-2 models every star particle as a single stellar population with a single age and metallicity, and tabulates all feedback event rates, luminosities and energies, mass-loss rates, and other quantities directly from stellar evolution models \citep[STARBURST99 v7.0;][]{Leitherer1999, Leitherer2014}, assuming a \citet{Kroupa2001} initial mass function for stars across $0.1 - 100 \Msun$.

Core-collapse supernovae, Ia supernovae, and stellar winds generate and deposit metals into surrounding gas cells.
FIRE-2 adopts the following models:
(1) for stellar winds, rates from STARBURST99 and yields from a compilation of \citet{VanDenHoek1997, Marigo2001, Izzard2004}; (2) for core-collapse supernovae, rates from STARBURST99 and nucleosynthetic yields from \citet{Nomoto2006}; (3) for Ia supernovae, rates from \citet{Mannucci2006} and yields from \citet{Iwamoto1999}.
FIRE-2 initializes abundances in gas (typically at $z \approx 99$) for all elements $i$ (beyond H and He) to a floor of $\left[\rm M_i / \rm H \right] \approx -4$, to prevent numerical problems in cooling.
All simulations in this data release (except the \textit{Massive Halo} suite) include an explicit model for unresolved turbulent diffusion of metals in gas \citep{Hopkins2017, Su2017, Escala2018}.

For more details on the physics and numerics of the FIRE-2 simulations, see \citet{Hopkins2015} for the \textsc{Gizmo} simulation code, \citet{Hopkins2018b} for the FIRE-2 physics model, \citet{Hopkins2018a} for more details on modeling mechanical feedback, and \citet{Hopkins2020a} for more details on modeling radiative feedback.

\subsection{Physics not modeled}
\label{sec:notmodel}

We release the FIRE-2 simulations that include the base set of FIRE-2 physics, as described above.
These simulations do \textit{not} include any additional physics, including the optional models in \citet{Hopkins2018b}.
Specifically, the simulations that we release:
\begin{itemize}
\item do not include \ac{MHD} or anisotropic conduction and viscosity; recent implementations in FIRE-2 suggest that these processes do not significantly change galaxy-wide properties \citep{Su2017, Hopkins2020b}.
\item do not model self-consistent cosmic-ray injection and transport, beyond assuming a fixed heating rate from cosmic rates in dense gas \citep[see][]{Chan2019, Ji2020, Hopkins2021b}
\item do not model self-consistent creation and destruction of dust, beyond simply assuming that dust traces gas-phase metallicity \citep[see][]{Choban2022}
\item use the LEBRON method to model radiative transfer in the optically thin limit (beyond the local gas kernel); they do not model radiation hydrodynamics via methods such as flux-limited diffusion or M1, though these approaches are unlikely to change galaxy-wide properties significantly \citep[see][]{Hopkins2020a}
\item only the \textit{Massive Halo} suite models the growth of super-massive black holes, and no simulation models feedback from an \ac{AGN} \citep[see][]{Wellons2022, MercedesFeliz2023}.
\end{itemize}
In many cases, these additional models remain under active development and exploration within the FIRE collaboration, and we anticipate including simulations that model them in future data releases.
We caution users about interpreting properties that may be sensitive to these physical processes: for example, the lack of \ac{MHD} will underestimate small-scale magnetic pressure in the \ac{ISM}, which could bias the properties of a structure like a \ac{GMC}.
See Section~\ref{sec:caveat} for more discussion.

All of the simulations that we release used the same FIRE-2 physics model (with minor variations as we describe).
While this provides a self-consistent suite, it does not allow a user to explore the effects of different astrophysical models or model parameters.
That said, we released the cosmological initial conditions and \textsc{Gizmo} configuration and parameter files for nearly all of these simulations, and a version of the \textsc{Gizmo} source code is publicly available.
Therefore, although the code including the full FIRE-2 physics is not presently in the public domain, users have access to tools necessary to (re)run simulations with model variants, including restarting simulations from the released snapshots.

\subsection{Zoom-in method}
\label{sec:zoom}

All FIRE-2 cosmological simulations zoom in on a selected region at high resolution, embedded within a lower-resolution cosmological background \citep[see][]{Onorbe2014}.
We first ran low-resolution dark-matter-only simulations within uniform-resolution cosmological boxes, then we selected regions of interest at $z\!=\!0$ for the \textit{Core} suite, $z\!=\!1$ for the \textit{Massive Halo} suite or $z\!=\!5$ for the \textit{High Redshift} suite.
We then chose a spherical volume centered on one halo (or a pair of halos) of interest.
For most FIRE-2 simulations, this region extends $4 - 8 \Rthm$ around the halo(s), where $\Rthm$ is the radius within which the mean density of the halo is 200 times the mean matter density of the universe.
We then traced the particles in this region back to $z \approx 99$ and regenerated the encompassing convex hull at high resolution using \textsc{MUSIC} \citep{HahnAbel2011}.
We resimulated the zoom-in region at high resolution, including dark matter, gas, and star formation, while the lower-resolution cosmological box that encompasses it contains only dark matter at low resolution.
As a result of using a convex hull to set/encompass the initial-condition volume (to help ensure its regularity and smoothness), the geometry of the zoom-in region at lower redshifts can be irregular and nonspherical.
By design, the primary halo(s) in each zoom-in region have zero contamination from low-resolution dark matter out to at least $\Rthm$ and typically much farther.

Each cosmological simulation zoom region is typically one to a few Mpc in size.
Except for the \textit{ELVIS on FIRE} simulations, we centered each zoom-in region on a single primary halo that we chose to be cosmologically isolated from halos of similar or greater mass, purely to limit computational cost.
(\citealt{Onorbe2014} showed via dark-matter-only simulations that the Lagrangian volume of the initial conditions of a halo does not bias its properties at $z\!=\!0$, though the effects of the initial conditions on galaxy properties in baryonic simulations remain less explored.)
Thus, an important caveat is that these simulations do not fairly sample the full range of cosmological environments.
For example, they do not sample the densest regions that a halo can inhabit, and there are no satellites of massive galaxy groups or clusters, nor ``splashback'' galaxies that ever orbited within them.
Furthermore, these simulations do not sample the lowest-density regions that probe the typical \ac{IGM}.

We chose most primary halos at particular mass scales, for example, $\Mthm(z\!=\!0) \sim 10^{9}, 10^{10}, 10^{11}, 10^{12} \Msun$ for the \textit{Core} suite, so the primary halos/galaxies in these simulations do not fairly sample the full halo/galaxy mass function.
In particular, we chose these systems initially based on their final dark-matter halo mass, so while the selection function of halo masses is well defined, the selection function of galaxy stellar masses is not, because of scatter in the relation between galaxy stellar mass and halo mass.
So, a set of primary galaxies at a given stellar mass does not necessarily sample the full range of halo masses that could form such galaxies.

Analyzing zoom-in simulations is different than analyzing a larger-volume, uniform-resolution cosmological simulation, like Illustris or EAGLE.
In particular, while each zoom-in simulation contains particles across the entire cosmological volume (typically $86 - 172 \Mpc$ along each spatial dimension), the volume outside of the zoom-in region contains only low-resolution dark-matter particles.
Generally, a user should analyze only high-resolution particles (which is straightforward, because \textsc{Gizmo} stores low-resolution dark-matter particles as a separate particle type, see Section~\ref{sec:snapshot}) that are safely within the zoom-in region.
As a simulation progresses, the zoom-in region inevitably develops a boundary region at its edge that contains overlapping high- and low-resolution particles, so one must use caution in analyzing high-resolution particles near the edge of the zoom-in region.
To make this easier, our default \textsc{Rockstar} halo HDF5 files (see Section~\ref{sec:halo}) include the total mass of low-resolution dark-matter particles within each halo.
We recommend analyzing only galaxies within halos whose fraction of total mass in low-resolution particles is less than a few percent.

\subsection{Additional caveats and limitations}
\label{sec:caveat}

\subsubsection{High redshifts}

In addition to using the ultraviolet background from \citet{FaucherGiguere2009}, which reionizes the universe at $z \approx 10$ \citep[rather than at $z \sim 8$, as recent empirical constraints favor,][]{FaucherGiguere2020}, these FIRE-2 simulations inadvertently suffer from spurious heating from cosmic rays in neutral gas at temperatures $\lesssim \! 1000$\,K at $z \gtrsim 10$ (before reionization), as noted in \citet{Su2018} (footnote 3) and \citet{GarrisonKimmel2019b} (Sec 3.3.2).
This term models (spatially uniform) cosmic-ray heating in the interstellar medium of a galaxy, but the version of \textsc{Gizmo} used for these suites erroneously applied it to low-density gas at $z \gtrsim 10$, before reionization, when the \ac{IGM} was both neutral and cold (after the gas temperature significantly decoupled from the cosmic microwave background at $z \lesssim 100$).
At these redshifts, this heating term suppressed star formation in low-mass halos, although its net effects are largely degenerate with the too-early reionization model, and it has no effect after reionization begins.
At $z \leq 4$, we confirmed that this cosmic-ray heating significantly affects only galaxy properties at masses $\Mstar \lesssim 10^{5} \Msun$; it does not significantly affect more massive galaxies, other than slightly reducing the (small) population of stars forming at $z \gtrsim 10$.
In light of this, when we simulated the 4 lowest-mass galaxies (m09, m10q, m10v, m11b)
in the \textit{Core} suite \citep[see][]{Wheeler2019},
we used a version of \textsc{Gizmo} that fixed this error.
Thus, those 4 simulations do not suffer from this spurious cosmic-ray heating at $z \gtrsim 10$, but all other simulations do.

\textit{Given the combination of this spurious heating term at $z \gtrsim 10$ and the ultraviolet background model that reionizes at $z \approx 10$ (likely too early) in FIRE-2, we caution users about interpreting galaxy properties, such as star-formation histories, at $z \gtrsim 8$.}

\subsubsection{Numerical limitations in resolution}

The tables in Section~\ref{sec:simulations} list the spatial resolution (gravitational force softening and gas smoothing) and particle mass for each particle species in each simulation.
These can differ across simulations, which a user should bear in mind, especially if combining simulations.

In general, one should trust only a structure/feature on scales greater than a few times (and ideally much more) these spatial resolution lengths.
Similarly, one should trust only an object resolved with $\gtrsim 10$ (and ideally many more) particles, including \ac{GMC}s, star clusters, (sub)halos, or low-mass galaxies.

Furthermore, the spatial resolution for gas is fully adaptive.
While this provides high spatial resolution in dense gas, conversely it means that the simulations typically resolve low-density gas, such as in the circumgalactic or intergalactic medium, with only $\gtrsim 1 \kpc$ spatial resolution.

See \citet{Hopkins2018b} for a comprehensive discussion of resolution tests and considerations.

\subsubsection{Known tensions with observations}

We next describe some of the known tensions with observations.
We emphasize that this refers \textit{only} to the FIRE-2 simulations with the base physics that we release in DR1; in many cases, the additional physics not modeled (see Section~\ref{sec:notmodel}) can help to alleviate such tension.

One of the most important limitations of these FIRE-2 simulations is the lack of \ac{AGN} feedback.
This causes the massive galaxies in the \textit{Massive Halo} suite to form too many stars, in ultradense nuclear distribution at late times \citep{Cochrane2019, Wellons2020, AnglesAlcazar2021, Parsotan2021, MercedesFeliz2023}, which is why we simulated these galaxies only to $z\!=\!1$.
The lack of \ac{AGN} feedback also may cause the \ac{MW}-mass galaxies in the \textit{Core} suite to form overly massive bulges, on average, and play a role in their possibly elevated star-formation rates at late cosmic times, on average \citep[][]{Chan2022, Gandhi2022, Wellons2022}.

Similarly, compared to observationally constrained relations between galaxy stellar mass and halo mass at $z \approx 0$ \citep[for example][]{Behroozi2019}, nearly all of our \ac{MW}-mass galaxies in the \textit{Core} suite lie $\approx 1 \sigma$ above the mean relation, that is, have high stellar mass for their halo mass \citep{Hopkins2018b}.
However, given that the \ac{MW} and M31 appear to lie above the average relation as well \citep[for example][]{BlandHawthornGerhard2016}, this means that the FIRE-2 simulations provide better analogs to \ac{MW} and M31, specifically.
As \citet{Hopkins2023} show, the newer FIRE-3 model leads to \ac{MW}-mass halos with lower stellar mass, more in line with the mean relation observed.

While the morphologies and kinematics of these FIRE-2 galaxies near \ac{MW} masses ($\Mstar \gtrsim 10^{10} \Msun$) and at much lower masses ($\Mstar \lesssim 10^{7} \Msun$ broadly agree with observations \citep[for example][]{Wheeler2017}, at intermediate masses ($\Mstar\!\sim\!10^{8 - 10} \Msun$) the FIRE-2 galaxies are insufficiently ``disky'', that is, too dispersion-dominated, as compared with observations \citep{ElBadry2018a, ElBadry2018b, KadoFong2022}.
Related to this, nearly all FIRE-2 galaxies at $\Mstar\!\sim\!10^{7 - 10} \Msun$ have extended sizes; essentially none of them form a compact, baryon-dominated, high-density stellar distribution, as observed in some galaxies at these masses \citep{GarrisonKimmel2019a, Shen2022}.
Both of these tensions reflect the difficulty, common to most modern cosmological simulations, in reproducing the diversity of galaxy rotation curves (dynamical masses) and sizes \citep[see][for review]{Sales2022}, which in the case of FIRE-2 may arise from excess burstiness in star formation at these masses \citep{Emami2019, Emami2021}.

Overly large sizes can extend to fainter galaxies in FIRE-2 as well.
Within the \ac{MW}-mass galaxy simulations, the lower-mass satellite galaxies tend to have sizes that are marginally larger than observed in the \ac{LG} \citep{Shen2022}, which in this case is mostly numerical, given the limited resolution of such low-mass galaxies in the \ac{MW}-mass simulations.
Still, even the much more highly resolved isolated faint galaxies (like m09, m10q, m10v) have sizes that tend to be larger than observed in the \ac{LG} \citep{Wheeler2019, Sales2022}, although the small sample size for FIRE-2 in this regime limits robust comparisons.

Regarding elemental abundances in gas and stars, given the assumed rates and yields of supernovae and stellar winds, the FIRE-2 simulations tend to moderately overestimate $\alpha$-element abundances (like O, Ca, Mg) and moderately underestimate yields from Ia supernova, in particular Fe \citep[][]{Escala2018, Hopkins2020b, Gandhi2022}, leading in particular to high normalizations in ratios like [$\alpha$/Fe].
Furthermore, while more massive galaxies in FIRE-2 show overall good agreement with the observed relation between stellar or gas metallicity and galaxy mass, galaxies at $\Mstar \lesssim 10^7 \Msun$, especially into the ultrafaint regime, have systematically lower [Fe/H] than observed by $0.2-0.5$ dex \citep{Wetzel2016, Escala2018, Wheeler2019, Hopkins2020b, Muley2021}.

\begin{deluxetable*}{ccccccccccccc}
\tablecaption{
\textit{Core} suite of 23 primary galaxies/halos across 20 different simulations to $z\!=\!0$;
we release 39 full snapshots across $z\!=\!0\!-\!10$.
Each cosmological simulation zooms in on a single isolated halo, except the last set (\textit{ELVIS on FIRE} suite) for which each simulation zooms in on a Local Group-like \ac{MW}+M31-mass pair (Romeo \& Juliet, Thelma \& Louise, Romulus \& Remus).
We simulated m09, m10q, m10v, and m11b without spurious cosmic-ray heating at $z \gtrsim 10$ (see Section~\ref{sec:core}).
}
\tablehead{
\colhead{name} & \colhead{$M_{\rm 200m}$} & \colhead{$R_{\rm 200m}$} & \colhead{$M_{\rm star,90}$} & \colhead{$m_{\rm baryon}$} & \colhead{$m_{\rm dm}$} & \colhead{$\epsilon_{\rm gas, min}$} & \colhead{$\epsilon_{\rm star}$} & \colhead{$\epsilon_{\rm dm}$} & \colhead{$N_{\rm dm}$} & \colhead{size} & \colhead{cosmology} & \colhead{reference} \\
& \colhead{[$\Msun$]} & \colhead{[kpc]} &  \colhead{[$\Msun$]} & \colhead{[$\Msun$]} & \colhead{[$\Msun$]} & \colhead{[pc]} & \colhead{[pc]} & \colhead{[pc]} & & \colhead{[GB]} & &
}
\startdata
m09 & $2.60 \times 10^9$ & 44 & $1.5 \times 10^4$ & 32 & 160 & 0.4 & 0.7 & 14 & $2.31 \times 10^8$ & 17 & N & \citetalias{Wheeler2019} \\
\hline
m10q & $8.23 \times 10^9$ & 64 & $4.8 \times 10^6$ & 33 & 160 & 0.4 & 0.7 & 14 & $1.24 \times 10^8$ & 11 & A & \citetalias{Wheeler2019} \\
m10v & $1.08 \times 10^{10}$ & 70 & $3.1 \times 10^5$ & 33 & 160 & 0.1 & 0.7 & 14 & $2.94 \times 10^8$ & 25 & A & \citetalias{Wheeler2019} \\
\hline
m11b & $4.65 \times 10^{10}$ & 114 & $4.5 \times 10^7$ & 2100 & 10,000 & 1.0 & 2.6 & 26 & $2.24 \times 10^7$ & 3.1 & A & \citetalias{Chan2018} \\
m11i & $7.77 \times 10^{10}$ & 133 & $9.2 \times 10^8$ & 7100 & 39,000 & 1.0 & 4.0 & 40 & $4.59 \times 10^6$ & 0.9 & P & \citetalias{ElBadry2018a} \\
m11q & $1.63 \times 10^{11}$ & 174 & $3.7 \times 10^8$ & 880 & 4400 & 0.5 & 2.0 & 20 & $1.30 \times 10^8$ & 11 & A & \citetalias{Hopkins2018b} \\
m11e & $1.68 \times 10^{11}$ & 171 & $1.4 \times 10^9$ & 7100 & 39,000 & 1.0 & 4.0 & 40 & $1.12 \times 10^7$ & 2.0 & P & \citetalias{ElBadry2018a} \\
m11h & $2.07 \times 10^{11}$ & 184 & $3.6 \times 10^9$ & 7100 & 39,000 & 1.0 & 4.0 & 40 & $1.58 \times 10^7$ & 2.8 & P & \citetalias{ElBadry2018a} \\
m11d & $3.23 \times 10^{11}$ & 213 & $3.9 \times 10^9$ & 7100 & 39,000 & 1.0 & 4.0 & 40 & $1.61 \times 10^7$ & 2.8 & P & \citetalias{ElBadry2018a} \\
\hline
m12z & $9.25 \times 10^{11}$ & 307 & $2.0 \times 10^{10}$ & 4200 & 21,000 & 0.4 & 3.2 & 33 & $1.27 \times 10^8$ & 14 & Z & \citetalias{GarrisonKimmel2019a} \\
m12w & $1.08 \times 10^{12}$ & 319 & $5.7 \times 10^{10}$ & 7100 & 39,000 & 1.0 & 4.0 & 40 & $7.33 \times 10^7$ & 8.4 & P & \citetalias{Samuel2020} \\
m12r & $1.10 \times 10^{12}$ & 321 & $1.7 \times 10^{10}$ & 7100 & 39,000 & 1.0 & 4.0 & 40 & $6.03 \times 10^7$ & 6.2 & P & \citetalias{Samuel2020} \\
m12i & $1.18 \times 10^{12}$ & 336 & $6.3 \times 10^{10}$ & 7100 & 35,000 & 1.0 & 4.0 & 40 & $7.05 \times 10^7$ & 7.2 & A & \citetalias{Wetzel2016} \\
m12c & $1.35 \times 10^{12}$ & 351 & $5.8 \times 10^{10}$ & 7100 & 35,000 & 1.0 & 4.0 & 40 & $1.51 \times 10^8$ & 16 & A & \citetalias{GarrisonKimmel2019a} \\
m12b & $1.43 \times 10^{12}$ & 358 & $8.5 \times 10^{10}$ & 7100 & 35,000 & 1.0 & 4.0 & 40 & $7.45 \times 10^7$ & 8.4 & A & \citetalias{GarrisonKimmel2019a} \\
m12m & $1.58 \times 10^{12}$ & 371 & $1.1 \times 10^{11}$ & 7100 & 35,000 & 1.0 & 4.0 & 40 & $1.41 \times 10^8$ & 17 & A & \citetalias{Hopkins2018b} \\
m12f & $1.71 \times 10^{12}$ & 380 & $7.9 \times 10^{10}$ & 7100 & 35,000 & 1.0 & 4.0 & 40 & $9.62 \times 10^7$ & 9.1 & A & \citetalias{GarrisonKimmel2017} \\
\hline
Juliet & $1.10 \times 10^{12}$ & 321 & $3.8 \times 10^{10}$ & 3500 & 19,000 & 0.7 & 4.4 & 32 & $3.06 \times 10^8$ & 33 & P & \citetalias{GarrisonKimmel2019a} \\
Romeo & $1.32 \times 10^{12}$ & 341 & $6.6 \times 10^{10}$ & 3500 & 19,000 & 0.7 & 4.4 & 32 & $3.06 \times 10^8$ & 33 & P & \citetalias{GarrisonKimmel2019a} \\
Louise & $1.15 \times 10^{12}$ & 333 & $2.6 \times 10^{10}$ & 4000 & 20,000 & 0.4 & 2.7 & 31 & $3.80 \times 10^8$ & 42 & E & \citetalias{GarrisonKimmel2019a} \\
Thelma & $1.43 \times 10^{12}$ & 358 & $7.1 \times 10^{10}$ & 4000 & 20,000 & 0.4 & 2.7 & 31 & $3.80 \times 10^8$ & 42 & E & \citetalias{GarrisonKimmel2019a} \\
Remus & $1.22 \times 10^{12}$ & 339 & $4.6 \times 10^{10}$ & 4000 & 20,000 & 0.4 & 3.8 & 31 & $2.97 \times 10^8$ & 33 & E & \citetalias{GarrisonKimmel2019b} \\
Romulus & $2.08 \times 10^{12}$ & 406 & $9.1 \times 10^{10}$ & 4000 & 20,000 & 0.4 & 3.8 & 31 & $2.97 \times 10^8$ & 33 & E & \citetalias{GarrisonKimmel2019b} \\
\enddata
\tablecomments{
We list the following properties for each galaxy/halo at $z\!=\!0$. \\
name: this generally indicates the (log) halo mass, to order of magnitude. \\
$M_{\rm 200m}$ and $R_{\rm 200m}$: total mass and spherical radius within which the mean density is $200 \times$ the matter density of the universe. \\
$M_{\rm star,90}$: stellar mass within a spherical radius that encloses 90\% of the stellar mass within 20 kpc. \\
$m_{\rm baryon}$ and $m_{\rm dm}$: initial masses of baryonic (gas or star) and dark-matter particles; gas cells can be up to 3 times more massive than this, because they gain mass from stellar ejecta/winds; for star particles this represents the typical mass at formation, but because of stellar mass loss, the typical star particle is $\approx 30\%$ smaller than this. \\
$\epsilon_{\rm gas, min}$: minimum adaptive force softening (Plummer equivalent) for gas cells (equals the hydrodynamic smoothing kernel). \\
$\epsilon_{\rm star}$ and $\epsilon_{\rm dm}$: force softening (Plummer equivalent) for star and dark-matter particles. \\
$N_{\rm dm}$: number of high-resolution dark-matter particles in the zoom-in region; the total number of particles (including gas and stars) is approximately twice this. \\
size: total size in GB of each simulation snapshot (for some simulations, a snapshot is stored across multiple file blocks). \\
Cosmology: cosmological parameters used in the simulation, as follows: \\
A (`AGORA': $\Omega_{\rm m} = 0.272, \Omega_{\rm \Lambda} = 0.728, \Omega_{\rm b} = 0.0455, h = 0.702, \sigma_{\rm 8} = 0.807,  n_{\rm s} = 0.961$); \\
P (`Planck': $\Omega_{\rm m} = 0.31, \Omega_{\rm \Lambda} = 0.69, \Omega_{\rm b} = 0.048, h = 0.68, \sigma_{\rm 8} = 0.82, n_{\rm s} = 0.97$); \\
N ($\Omega_{\rm m} = 0.266, \Omega_{\rm \Lambda} = 0.734, \Omega_{\rm b} = 0.044, h = 0.71, \sigma_{\rm 8} = 0.801,  n_{\rm s} = 0.963$); \\
E ($\Omega_{\rm m} = 0.266, \Omega_{\rm \Lambda} = 0.734, \Omega_{\rm b} = 0.0449, h = 0.71, \sigma_{\rm 8} = 0.801,  n_{\rm s} = 0.963$). \\
Z ($\Omega_{\rm m} = 0.2821, \Omega_{\rm \Lambda} = 0.7179, \Omega_{\rm b} = 0.0461, h = 0.697, \sigma_{\rm 8} = 0.817,  n_{\rm s} = 0.9646$); \\
Reference: published article that introduced the simulation at this resolution. We request any user of a given simulation to cite this article:
\citealt{Wetzel2016} \citepalias{Wetzel2016};
\citealt{GarrisonKimmel2017} \citepalias{GarrisonKimmel2017};
\citealt{Hopkins2018b} \citepalias{Hopkins2018b};
\citealt{Chan2018} \citepalias{Chan2018};
\citealt{ElBadry2018a} \citepalias{ElBadry2018a};
\citealt{Wheeler2019} \citepalias{Wheeler2019};
\citealt{GarrisonKimmel2019a} \citepalias{GarrisonKimmel2019a};
\citealt{GarrisonKimmel2019b} \citepalias{GarrisonKimmel2019b};
\citealt{Samuel2020} \citepalias{Samuel2020}.
}
\label{tab:core}
\end{deluxetable*}

\begin{deluxetable*}{ccccccccccccc}
\tablecaption{
\textit{Massive Halo} suite of 4 primary galaxies/halos simulated to $z\!=\!1$;
we release 19 full snapshots across $z\!=\!1\!-\!10$.
}
\tablehead{
\colhead{name} & \colhead{$M_{\rm vir}$} & \colhead{$R_{\rm vir}$} & \colhead{$M_{\rm star}$} & \colhead{$m_{\rm baryon}$} & \colhead{$m_{\rm dm}$} & \colhead{$\epsilon_{\rm gas, min}$} & \colhead{$\epsilon_{\rm star}$} & \colhead{$\epsilon_{\rm dm}$} & \colhead{$N_{\rm dm}$} & \colhead{size} & \colhead{cosmology} & \colhead{reference} \\
& \colhead{[$\Msun$]} & \colhead{[kpc]} &  \colhead{[$\Msun$]} & \colhead{[$\Msun$]} & \colhead{[$\Msun$]} & \colhead{[pc]} & \colhead{[pc]} & \colhead{[pc]} & & \colhead{[GB]} & &
}
\startdata
A1 & $3.92 \times 10^{12}$ & 247 & $2.75 \times 10^{11}$ & $3.3\times10^4$ & $1.7\times10^5$ & 0.7 & 7 & 57 & $3.52 \times 10^7$ & 7.3 & Z & \citetalias{AnglesAlcazar2017b} \\
A2 & $7.75 \times 10^{12}$ & 310 & $4.10 \times 10^{11}$ & $3.3\times10^4$ & $1.7\times10^5$ & 0.7 & 7 & 57 & $1.13 \times 10^8$ & 23 & Z & \citetalias{AnglesAlcazar2017b} \\
A4 & $4.54 \times 10^{12}$ & 260 & $2.34 \times 10^{11}$ & $3.3\times10^4$ & $1.7\times10^5$ & 0.7 & 7 & 57 & $6.44 \times 10^7$ & 13 & Z & \citetalias{AnglesAlcazar2017b} \\
A8 & $1.27 \times 10^{13}$ & 359 & $5.36 \times 10^{11}$ & $3.3\times10^4$ & $1.7\times10^5$ & 0.7 & 7 & 57 & $1.42 \times 10^8$ & 29 & Z & \citetalias{AnglesAlcazar2017b} \\
\hline
\enddata
\tablecomments{
Properties are as in Table~\ref{tab:core}, with the following exceptions.
We list galaxy/halo properties at $z\!=\!1$.
$M_{\rm vir}$ and $R_{\rm vir}$ refer to the evolving virial overdensity definition for a halo from \citet{BryanNorman1998}, and $M_{\rm star}$ is the total stellar mass within $0.1\,R_{\rm vir}$.
We request any user of these simulations to cite \citealt{AnglesAlcazar2017b} \citepalias{AnglesAlcazar2017b}.
}
\label{tab:massive}
\end{deluxetable*}

\begin{deluxetable*}{ccccccccccccc}
\tablecaption{
\textit{High Redshift} suite of 22 primary galaxies/halos simulated to $z\!=\!5$;
we release 11 full snapshots across $z\!=\!5\!-\!10$.
}
\tablehead{
\colhead{name} & \colhead{$M_{\rm vir}$} & \colhead{$R_{\rm vir}$} & \colhead{$M_{\rm star}$} & \colhead{$m_{\rm baryon}$} & \colhead{$m_{\rm dm}$} & \colhead{$\epsilon_{\rm gas, min}$} & \colhead{$\epsilon_{\rm star}$} & \colhead{$\epsilon_{\rm dm}$} & \colhead{$N_{\rm dm}$} & \colhead{size} & \colhead{cosmology} & \colhead{reference} \\
& \colhead{[$\Msun$]} & \colhead{[kpc]} &  \colhead{[$\Msun$]} & \colhead{[$\Msun$]} & \colhead{[$\Msun$]} & \colhead{[pc]} & \colhead{[pc]} & \colhead{[pc]} & & \colhead{[GB]} & &
}
\startdata
z5m12b & $8.7\times10^{11}$ & 51.2 & $2.6\times10^{10}$ & 7100 & $3.9\times10^4$ & 0.42 & 2.1 & 42 & $3.5\times10^7$ & 8.3 & P & \citetalias{Ma2018a} \\
z5m12c & $7.9\times10^{11}$ & 49.5 & $1.8\times10^{10}$ & 7100 & $3.9\times10^4$ & 0.42 & 2.1 & 42 & $3.3\times10^7$ & 8.6 & P & \citetalias{Ma2019} \\
z5m12d & $5.7\times10^{11}$ & 44.5 & $1.2\times10^{10}$ & 7100 & $3.9\times10^4$ & 0.42 & 2.1 & 42 & $2.3\times10^7$ & 7.6 & P & \citetalias{Ma2019} \\
z5m12e & $5.0\times10^{11}$ & 42.6 & $1.4\times10^{10}$ & 7100 & $3.9\times10^4$ & 0.42 & 2.1 & 42 & $1.9\times10^7$ & 6.3 & P & \citetalias{Ma2019} \\
z5m12a & $4.5\times10^{11}$ & 41.1 & $5.4\times10^9$ & 7100 & $3.9\times10^4$ & 0.42 & 2.1 & 42 & $1.7\times10^7$ & 4.8 & P & \citetalias{Ma2018a} \\
z5m11f & $3.1\times10^{11}$ & 36.4 & $4.7\times10^9$ & 7100 & $3.9\times10^4$ & 0.42 & 2.1 & 42 & $1.2\times10^7$ & 4.0 & P & \citetalias{Ma2019} \\
z5m11e & $2.5\times10^{11}$ & 33.6 & $2.5\times10^9$ & 7100 & $3.9\times10^4$ & 0.42 & 2.1 & 42 & $9.6\times10^6$ & 2.6 & P & \citetalias{Ma2018a} \\
z5m11g & $2.0\times10^{11}$ & 31.2 & $1.9\times10^9$ & 7100 & $3.9\times10^4$ & 0.42 & 2.1 & 42 & $7.7\times10^6$ & 2.6 & P & \citetalias{Ma2019} \\
z5m11d & $1.4\times10^{11}$ & 27.5 & $1.6\times10^9$ & 7100 & $3.9\times10^4$ & 0.42 & 2.1 & 42 & $4.8\times10^6$ & 1.7 & P & \citetalias{Ma2018a} \\
z5m11h & $1.0\times10^{11}$ & 24.9 & $1.6\times10^9$ & 7100 & $3.9\times10^4$ & 0.42 & 2.1 & 42 & $3.8\times10^6$ & 1.6 & P & \citetalias{Ma2019} \\
z5m11c & $7.6\times10^{10}$ & 22.7 & $9.5\times10^8$ & 891 & 4900 & 0.28 & 1.4 & 21 & $2.0\times10^7$ & 9.1 & P & \citetalias{Ma2020b} \\
z5m11i & $5.2\times10^{10}$ & 20.0 & $2.8\times10^8$ & 891 & 4900 & 0.28 & 1.4 & 21 & $1.3\times10^7$ & 6.3 & P & \citetalias{Ma2020b} \\
z5m11b & $4.0\times10^{10}$ & 18.3 & $1.7\times10^8$ & 891 & 4900 & 0.28 & 1.4 & 21 & $1.1\times10^7$ & 5.2 & P & \citetalias{Ma2018a} \\
z5m11a & $4.2\times10^{10}$ & 18.6 & $1.2\times10^8$ & 954 & 5200 & 0.28 & 1.4 & 21 & $1.0\times10^7$ & 5.3 & P & \citetalias{Ma2018a} \\
z5m10f & $3.3\times10^{10}$ & 17.2 & $1.6\times10^8$ & 954 & 5200 & 0.28 & 1.4 & 21 & $8.4\times10^6$ & 3.8 & P & \citetalias{Ma2018a} \\
z5m10e & $2.6\times10^{10}$ & 15.8 & $3.9\times10^7$ & 954 & 5200 & 0.28 & 1.4 & 21 & $7.5\times10^6$ & 3.3 & P & \citetalias{Ma2018a} \\
z5m10d & $1.9\times10^{10}$ & 14.2 & $4.8\times10^7$ & 954 & 5200 & 0.28 & 1.4 & 21 & $4.7\times10^6$ & 2.2 & P & \citetalias{Ma2018a} \\
z5m10c & $1.3\times10^{10}$ & 12.7 & $5.6\times10^7$ & 954 & 5200 & 0.28 & 1.4 & 21 & $3.2\times10^6$ & 1.5 & P & \citetalias{Ma2018a} \\
z5m10b & $1.2\times10^{10}$ & 12.4 & $3.4\times10^7$ & 954 & 5200 & 0.28 & 1.4 & 21 & $3.1\times10^6$ & 1.4 & P & \citetalias{Ma2018a} \\
z5m10a & $6.6\times10^9$ & 10.0 & $1.5\times10^7$ & 119 & 650 & 0.14 & 0.7 & 10 & $1.2\times10^7$ & 9.1 & P & \citetalias{Ma2020b} \\
z5m09b & $3.9\times10^9$ & 8.4 & $2.8\times10^6$ & 119 & 650 & 0.14 & 0.7 & 10 & $7.5\times10^6$ & 3.8 & P & \citetalias{Ma2018a} \\
z5m09a & $2.4\times10^9$ & 7.1 & $1.6\times10^6$ & 119 & 650 & 0.14 & 0.7 & 10 & $4.8\times10^6$ & 2.3 & P & \citetalias{Ma2018a} \\
\hline
\enddata
\tablecomments{
Properties are as in Table~\ref{tab:core}, with the following exceptions.
We list galaxy/halo properties at $z\!=\!5$.
$M_{\rm vir}$ and $R_{\rm vir}$ refer to the evolving virial overdensity definition for a halo from \citet{BryanNorman1998}, and $M_{\rm star}$ is the total stellar mass within $R_{\rm vir}$.
We request any user of these simulations to cite \citealt{Ma2018a}, \citealt{Ma2019}, and/or \citealt{Ma2020b}.
}
\label{tab:highz}
\end{deluxetable*}

\section{FIRE-2 suites of simulations}
\label{sec:simulations}

\subsection{Core suite to $z = 0$}
\label{sec:core}

Table~\ref{tab:core} lists the \textit{Core} suite of FIRE-2 simulations run to $z\!=\!0$, including the properties of each primary halo/galaxy at $z\!=\!0$.
We release 39 full snapshots across $z\!=\!0-10$.
Specifically, we release 19 snapshots across $z\!=\!1-10$ spaced every $\Delta z\!=\!0.5$, 9 snapshots across $z\!=\!0.1-1$ spaced every $\Delta z\!=\!0.1$, and 11 snapshots spaced every $\Delta t \approx 2.2 \Myr$ just prior to and including $z\!=\!0$.
Table~\ref{tab:core} also lists the published article that introduced each simulation at the stated resolution.
We request anyone who uses a given simulation to cite its relevant publication.

Except for the last set, we name the simulations according to the (log) mass of the primary host halo at $z\!=\!0$ (the letter in the name is arbitrary).
We selected these halos at $z\!=\!0$ based on their dark-matter halo mass, and an additional isolation criterion of having no neighboring halos of similar mass (typically $\gtrsim 30\%$) within at least (typically) $\approx 5 \, \Rthm$, motivated purely by limiting computational cost.

The bottom 2 sets in Table~\ref{tab:core} represent our suite of \ac{MW}/M31-mass galaxies.
Simulations named m12* (except m12z) we generated as part of the \textit{Latte} suite \citep[introduced in][]{Wetzel2016} of halos with $\Mthm(z\!=\!0) = 1\!-\!2 \times 10^{12} \Msun$.
We reemphasize that their selection was agnostic to any halo properties beyond mass, including formation history, concentration, spin, or satellite/subhalo population.
m12z is similar, although at slightly lower mass and better resolution.
Simulations in the last set are part of the \textit{ELVIS on FIRE} suite of \ac{LG}-like \ac{MW}+M31-mass pairs (Romeo \& Juliet, Thelma \& Louise, Romulus \& Remus).
Their selection at $z\!=\!0$ is \citep{GarrisonKimmel2019a}:
(1) two neighboring halos, each with a mass $\Mthm\!=\!1\!-\!3 \times 10^{12} \Msun$,
(2) total pair mass of $\Mthm\!=\!2\!-\!5 \times 10^{12} \Msun$,
(3) halo center separation of $600 - 1000 \kpc$,
(4) relative halo radial velocity $v_{\rm rad} < 0 \kms$,
and (5) no other massive halo within 2.8 Mpc of either host center.
These criteria do not constrain the larger-scale environment around these halos.

Given that users may be interested in comparing these simulations against the \ac{MW} (and M31), we note that, among the \textit{Latte} suite, m12i, m12f, m12m, m12b are probably the most similar to the \ac{MW} across a range of properties: \citet{Sanderson2020} showed that m12i, m12f, and m12m have broadly similar stellar masses, scale radii, scale heights, and gas fractions as the \ac{MW}.
Among the \textit{ELVIS on FIRE} suite, the thinnest, most Milky-Way-like disks are Romeo, Romulus, and Remus.
Relative to the \textit{Latte} suite, the galaxies in the \textit{ELVIS on FIRE} suite tend to form stars earlier \citep{Santistevan2020}, form larger disks \citep{Bellardini2022}, and their disks start to form/settle earlier \citep{Yu2021, McCluskey2023}.
In particular, Romeo is the earliest-forming galaxy/disk in the suite.
This may be relevant given that the \ac{MW}'s disk shows evidence for early formation \citep[for example][]{BelokurovKravtsov2022, Conroy2022}.

As Table~\ref{tab:core} shows, these simulations used similar but slightly different assumed cosmologies (generally for comparison with specific previous studies), encompassing the ranges: $\Omega_{\rm m} = 0.266 - 0.31$, $\Omega_{\rm \Lambda} = 0.69 - 0.734$, $\Omega_{\rm b} = 0.044 - 0.48$, $\sigma_{\rm 8} = 0.801 - 0.82$, $n_{\rm s} = 0.961 - 0.97$, and $w = -1$, generally consistent with \citet{Planck2020b}.
Some simulations used the cosmological box from the AGORA project \citep{Kim2014}.
Differences in growth histories from differing cosmological parameters are generally small compared with halo-to-halo variations.

\subsection{Massive Halo suite to $z = 1$}
\label{sec:massive}

Table~\ref{tab:massive} lists the \textit{Massive Halo} suite of FIRE-2 simulations run to $z\!=\!1$, including the properties of each primary halo/galaxy at $z\!=\!1$.
We release 19 full snapshots across $z\!=\!1-10$, spaced every $\Delta z\!=\!0.5$.
We request any user of these simulations to cite \citet{AnglesAlcazar2017b}.

We selected these halos from the A-series of the FIRE-1 \textit{MassiveFIRE} suite \citep{Feldmann2016, Feldmann2017} to cover a range of formation histories for halo mass $M_{\rm vir} \approx 10^{12.5}\Msun$ at $z\!=\!2$.
Refer to \citet{Feldmann2016, Feldmann2017} regarding the selection strategy and halo growth histories of the \textit{MassiveFIRE} simulations.

In addition to being re-simulated with the FIRE-2 model, these \textit{Massive Halo} simulations include a model for the growth of massive black holes, based on gravitational torques between the stellar and gas components \citep{Hopkins2011sf, AnglesAlcazar2017a}.
However, these simulations do not include \ac{AGN} feedback from black holes, so they form overly massive galaxies with ultradense nuclear stellar distributions at late times (see Section~\ref{sec:caveat}).
This is the key reason we simulated these galaxies only to $z\!=\!1$.

Unlike all other FIRE-2 simulations in this data release, these \textit{Massive Halo} simulations do not include a model for subgrid turbulent diffusion of metals in gas.

\subsection{High Redshift suite to $z = 5$}
\label{sec:highz}

Table~\ref{tab:highz} lists the \textit{High Redshift} suite of FIRE-2 simulations run to $z\!=\!5$, including the properties of each primary halo/galaxy at $z\!=\!5$.
We release 11 full snapshots across $z\!=\!5\!-\!10$, spaced every $\Delta z\!=\!0.5$.
We request any users to cite \citealt{Ma2018a}, \citealt{Ma2019}, and/or \citealt{Ma2020b}.

We generated these simulations for studying galaxies at the epoch of reionization \citep[see][]{Ma2018a, Ma2018b, Ma2019, Ma2020b}.
We selected these halos across a mass range of $M_{\rm vir} \approx 10^9\!-\!10^{12} \Msun$ at $z\!=\!5$ from cosmological volumes of size $(11\,{\rm Mpc})^3$ and $(43\,{\rm Mpc})^3$.
Including both the primary galaxy and all lower-mass (satellite) galaxies within each zoom-in region, this \textit{High Redshift} suite contains about 2000 resolved galaxies at $z\!=\!5$.

\begin{figure*}
\centering
\includegraphics[width = 0.95 \linewidth]{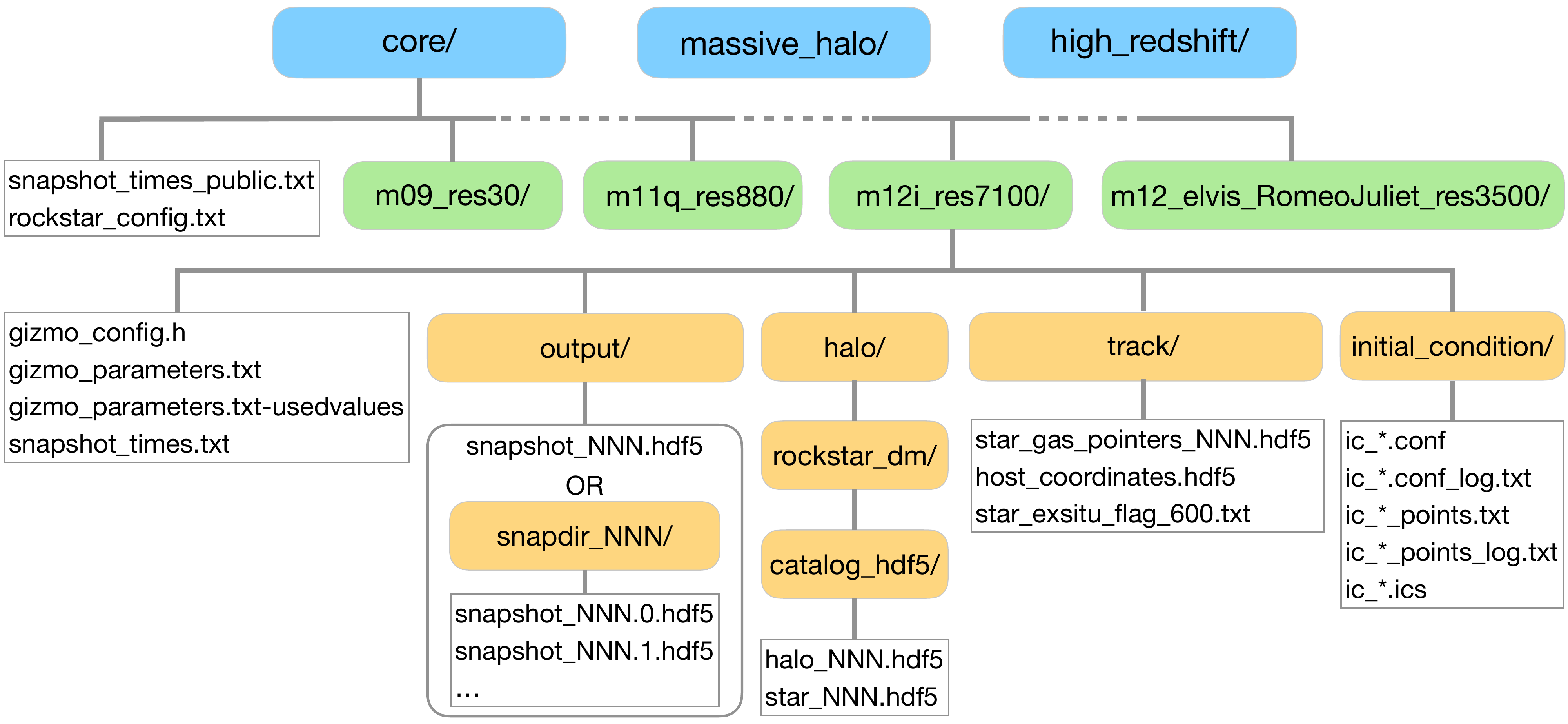}
\caption{
Schematic diagram of FIRE-2 simulation data, showing the names of directories and files as available via FlatHUB at \href{http://flathub.flatironinstitute.org/fire}{flathub.flatironinstitute.org/fire}.
This data release includes 3 suites of simulations, each within its own directory (top row in blue): the \textit{Core} suite to $z\!=\!0$, the \textit{Massive Halo} suite to $z\!=\!1$, and the \textit{High Redshift} suite to $z\!=\!5$.
Each suite contains a directory for each simulation (second row in green).
This diagram shows the contents of the m12i simulation from the \textit{Core} suite as an example, where \texttt{NNN} represents each snapshot index and * represents a wild card in the file name.
All suites and simulations have similar data structure, with minor variations.
See Section~\ref{sec:data} for details.
}
\label{fig:diagram}
\end{figure*}

\section{Data structure and access}
\label{sec:data}

Figure~\ref{fig:diagram} shows a schematic diagram of the structure of the FIRE-2 simulation data.
Each simulation suite resides in its own top-level directory, which in turn contains one directory for each simulation. 
We describe the data contents of these simulations in detail below.

\subsection{Accessing data}
\label{sec:access}

The FIRE-2 simulations are available via the Flatiron Institute Data Exploration and Comparison Hub (FlatHUB), at the following website:
\href{http://flathub.flatironinstitute.org/fire}{flathub.flatironinstitute.org/fire}.
FlatHUB provides two ways to access the data.
First, using the website above, a user can click on the ``Browse" box to access each suite and simulation via the browser.
We recommend this method to browse the available data and download a small amount of it.
Second, the FlatHUB website above also provides a Globus ID for transferring via Globus.\footnote{
\href{https://app.globus.org}{app.globus.org}
}
We recommend using Globus especially when transferring a large amount of data.

\subsection{Simulation snapshots}
\label{sec:snapshot}

Within a given simulation directory, all simulation snapshots are in a directory named \texttt{output/}.
\textsc{Gizmo} stores each snapshot via HDF5 file(s).
For simulations with fewer particles, we store each snapshot as a single HDF5 file, named \texttt{snapshot\_NNN.hdf5}, while we split larger simulations into multiple HDF5 files within a directory named \texttt{snapdir\_NNN/}.

\texttt{NNN} is the snapshot index, which increases with time and ranges from 0 to 600.
For example, at integer redshifts the snapshot indices are: $z\!=\!0$ (600), $z\!=\!1$ (277), $z\!=\!2$ (172), $z\!=\!3$ (120), $z\!=\!4$ (88), $z\!=\!5$ (67), $z\!=\!6$ (52), $z\!=\!7$ (41), $z\!=\!8$ (33), $z\!=\!9$ (26), $z\!=\!10$ (20).
Each simulation directory contains a file named \texttt{snapshot\_times.txt} that lists the index, scale factor, redshift, age of the universe (in Gyr), lookback time (in Gyr), and time spacing since the previous snapshot (in Myr) of all snapshots written (up to 600).
Each suite at its top level also contains a file named \texttt{snapshot\_times\_public.txt} that lists only the snapshots that we publicly release.

Each snapshot contains 4 types of particle species: gas cells (stored as type 0), stars (stored as type 4), and dark matter at high resolution (stored as type 1), all of which exist only in the zoom-in region, as well as low-resolution dark matter (stored as type 2) that exists across the entire cosmological box.

Each snapshot file contains an HDF5 header that includes useful information about the simulation and the contents of the snapshot, including the number of particles of each species, units, cosmological parameters, and so on.
Two of the most important (for unit conversions below) are: the scale factor of the snapshot, $a$, stored in the header as \texttt{Time}, and the dimensionless Hubble parameter, $h$, stored in the header as \texttt{HubbleParam}.

Below we list all properties stored for each particle species, along with their units within the snapshot file.
However, \textit{we strongly encourage anyone to use one of the publicly available python reader/analysis packages that we list in Section~\ref{sec:analysis}, which automatically convert all quantities into more common and useful units}.
For more extensive documentation on the contents of snapshots, refer to the \href{http://www.tapir.caltech.edu/~phopkins/Site/GIZMO_files/gizmo_documentation.html}{\textsc{Gizmo} Users Guide} (see Section \ref{sec:documentation}).

Each \textsc{Gizmo} snapshot stores the following properties for all particles, with the following names and units.
Any quantities listed without units are dimensionless.
\begin{itemize}
\item \texttt{ParticleIDs} - indexing starts at 0
(not necessarily unique for star particles and gas cells, see below)
\item \texttt{ParticleChildIDsNumber} and \\ \texttt{ParticleIDGenerationNumber} - star particles and gas cells have these two additional IDs to track them uniquely.
As a gas cell inherits mass from stellar feedback, to ensure mass balancing \textsc{Gizmo} splits it into two if it exceeds 3 times its initial mass, which means that multiple gas cells and/or star particles can have the same \texttt{ParticleID} (a star particle inherits its IDs from its progenitor gas cell).
Thus, \textsc{Gizmo} stores these two additional IDs, initialized to 0 at the start of the simulation.
Each time a gas cell splits in two, one cell retains the same \texttt{ParticleChildIDsNumber}, the other cell gets \texttt{ParticleChildIDsNumber} $\mathrel{+}=\!2^\texttt{ParticleIDGenerationNumber}$.
Both cells then get \texttt{ParticleIDGenerationNumber} $\mathrel{+}=\!1$.
Because \textsc{Gizmo} stores \texttt{ParticleChildIDsNumber} as a 32-bit integer, this allows for a maximum of 30 splittings, then \texttt{ParticleChildIDsNumber} aliases back to 0 and is no longer unique.
\item \texttt{Coordinates} [$h^{-1} \kpc$ comoving] - 3-D positions; multiply by the scale factor $a$ to convert to physical position
\item \texttt{Velocities} [$\sqrt{a}$ km s$^{-1}$] - 3-D velocities; multiply by $\sqrt{a}$ to convert to physical/peculiar velocity
\item \texttt{Masses} [$10^{10} \, h^{-1} \Msun$] - multiply by $10^{10} \, h^{-1}$ to convert to $\Msun$
\item \texttt{Potential} [km$^2$ s$^{-2}$] - gravitational potential with arbitrary normalization (stored for most \ac{MW}-mass simulations)
\item \texttt{OStarNumber} - simulations in the \textit{Core} suite that used baryonic mass resolution $\approx 30 \Msun$ (m09, m10q, m10v) used stochastic sampling of massive stars ($> 8 \Msun$), so this indicates the number of such stars in a given star particle
\end{itemize}

Star particles and gas cells also store their elemental abundances:
\begin{itemize}
\item \texttt{Metallicity} - 11 elemental abundances, stored as linear mass fractions, with the following order/indices: H (0), He (1), C (2), N (3), O (4), Ne (5), Mg (6), Si (7), S (8), Ca (9), Fe (10)\footnote{
All simulations contain at least these standard 11 elemental abundances in an array of shape $N_{\rm particle} \times 11$. Some simulations also employed a set of models for $r$-process enrichment \citet[see][]{vandeVoort2015}, which we simply appended to the end of each particle's metallicity array. Several simulations, including some in the \text{Core} and all in the {High Redshift} suites, included 4 different models for $r$-process and thus have total shape $N_{\rm particle} \times 15$, while the \textit{Massive Halo} suite included 6 different models for $r$-process and thus have total shape $N_{\rm particle} \times 17$.
}
\end{itemize}

Star particles also store:
\begin{itemize}
\item \texttt{StellarFormationTime} - scale factor at formation
\end{itemize}

Gas cells also store:
\begin{itemize}
\item \texttt{Density} [$10^{10} \, h^{2} a^{-3} \Msun \kpc^{-3}$] - defined via the cell's mass and the cell's \texttt{SmoothingLength}
\item \texttt{InternalEnergy} [km$^2$ s$^{-2}$] - specific internal energy; use to compute temperature
\item \texttt{SmoothingLength} [$h^{-1} \kpc$ comoving] - full extent of the neighbor interaction kernel (radius of compact support)
\item \texttt{ElectronAbundance} - mean number of free electrons per hydrogen nucleus
\item \texttt{NeutralHydrogenAbundance} - fraction of hydrogen that is neutral
\item \texttt{StarFormationRate} [$\Msun \yr^{-1}$] - instantaneous rate of star formation
\end{itemize}

Black-hole particles (only modeled in the \textit{Massive Halo} suite) also store:
\begin{itemize}
\item \texttt{BH\_Mass} [$10^{10} \, h^{-1} \Msun$] - mass of the black hole (not necessarily the total mass of the particle, see below)
\item \texttt{BH\_Mass\_AlphaDisk} [$10^{10} \, h^{-1} \Msun$] - mass in the subgrid viscous accretion disk
\item \texttt{BH\_Mdot} [$10^{10} \Msun \yr^{-1}$] - instantaneous rate of accretion
\item \texttt{BH\_AccretionLength} [$h^{-1} \kpc$ comoving] - full extent of the neighbor accretion radius (kernel length)
\item \texttt{BH\_NProgs} - cumulative number of black holes that merged into this one (0 if none)
\end{itemize}

\subsection{\textsc{Gizmo} settings}

Each simulation directory contains the following files that specify the settings that \textsc{Gizmo} used when compiling and running the simulation: \texttt{gizmo\_config.h} lists the compile-time configuration settings, and \texttt{gizmo\_parameters.txt} lists the run-time parameters.

\subsection{Initial Conditions}

For each simulation, we include its cosmological initial conditions\footnote{
Also available at \href{http://www.tapir.caltech.edu/~phopkins/publicICs}{www.tapir.caltech.edu/$\sim$phopkins/publicICs}
}
in a directory named \texttt{initial\_condition/}.
This contains the MUSIC configuration files, which list the full cosmological parameters, and the initial-condition file at $z \approx 99$, named \texttt{*.ics}.

\subsection{Catalogs of (sub)halos and galaxies}
\label{sec:halo}

Each simulation includes a catalog of (sub)halos and their galaxies at each snapshot, within a directory named \texttt{halo/}.

For each simulation, we generated (sub)halo catalogs via \textsc{Rockstar} \citep{Behroozi2013a},
We provide these as our default and recommended galaxy/halo catalogs.
We used a slightly \href{https://bitbucket.org/awetzel/rockstar-galaxies}{modified version}\footnote{
\href{https://bitbucket.org/awetzel/rockstar-galaxies}{bitbucket.org/awetzel/rockstar-galaxies}
} of \textsc{Rockstar-Galaxies}\footnote{
\href{https://bitbucket.org/pbehroozi/rockstar-galaxies}{bitbucket.org/pbehroozi/rockstar-galaxies}
}, which is a version of \textsc{Rockstar} with support for multimass and multispecies particles.
We used the same \textsc{Rockstar} parameters for all simulations, and we provide the input configuration file that we used, named \texttt{rockstar\_config.txt}, within the top-level directory of each suite.

We ran \textsc{Rockstar-Galaxies} using \textit{only} dark-matter particles, because we found this led to better numerical stability, especially for subhalos.
We therefore place these files in a directory named \texttt{rockstar\_dm/}, to reinforce that we generated the halo catalogs using only dark-matter information.
Thus, any (sub)halo properties in the catalog are measured using only the dark-matter particles (ignoring stars and gas).
We then assigned star particles to these (sub)halos in post-processing, using \href{https://bitbucket.org/awetzel/halo_analysis}{\textsc{HaloAnalysis}}\footnote{
\href{https://bitbucket.org/awetzel/halo_analysis}{bitbucket.org/awetzel/halo\_analysis}
} \citep{Wetzel2020b}; for more details see \citet{Samuel2020}.
We store these (sub)halo catalogs in a converted HDF5 format, named \texttt{halo\_NNN.hdf5}, and corresponding galaxy stellar and star-particle information for each (sub)halo is in \texttt{star\_NNN.hdf5}, where \texttt{NNN} is the snapshot index, all within a directory named \texttt{catalog\_hdf5/}.
Appendix~\ref{sec:rockstar} lists the halo/galaxy properties in these HDF5 files.

We strongly recommend using these HDF5 halo/galaxy files.
For completeness, however, we also provide the ASCII text files, \texttt{out\_NNN.list}, that \textsc{Rockstar} directly outputs in a separate directory named \texttt{catalog/}; see the documentation from \textsc{Rockstar-Galaxies} regarding the contents of these files.

For the \textit{Massive Halo} and \textit{High Redshift} suites, we also generated (sub)halo catalogs using the Amiga Halo Finder \citep[\textsc{AHF};][]{Knollmann2009},
which reside in the directory named \texttt{AHF/} within \texttt{halo/}.
We ran \textsc{AHF} simultaneously on all particles, including dark matter, gas, and stars.
\textsc{AHF} uses an isodensity contour to identify a halo center, and we defined the halo boundary via a spherical overdensity with a virial radius given by the redshift-dependent virial overdensity definition of \citet{BryanNorman1998}.
The \textsc{AHF} catalogs are in simple text format and contain many properties for (sub)halos, including stellar and gaseous properties.
See the \textsc{AHF} file header for more information.

\subsection{Pointers to track stars and gas across snapshots}
\label{sec:pointer}

For the \textit{Core} suite to $z\!=\!0$, each simulation also contains, within a directory named \texttt{track/}, HDF5 files named \texttt{star\_gas\_pointers\_NNN.hdf5}.
Each file contains, for every star particle and gas cell at snapshot 600 ($z\!=\!0$), a pointer to its index in the particle or cell array at a previous snapshot \texttt{NNN}.
Therefore, one can use these pointers easily to track where a star particle or gas cell was in a previous snapshot, or between any two snapshots.
We generated these pointers, because one cannot simply use \texttt{ParticleIDs} alone to match/track particles, because multiple gas cells and/or star particles can have the same \texttt{ParticleIDs}.
Rather, one needs to use \texttt{ParticleIDs} plus
\texttt{ParticleChildIDsNumber} and \texttt{ParticleIDGenerationNumber} (see Section~\ref{sec:snapshot}).
Thus, the pointers in \texttt{star\_gas\_pointers\_NNN.hdf5} merely simplify this particle tracking for a user.
See the \href{https://bitbucket.org/awetzel/gizmo_analysis}{\textsc{GizmoAnalysis}} package (Section~\ref{sec:analysis}) for more details on using them.

\subsection{Identifying the primary galaxy/halo}
\label{sec:center}

We describe two methods to locate the primary galaxy/halo within each simulation, specifically, its position and velocity.

First, one can use the \textsc{Rockstar} (or \textsc{AHF}) halo/galaxy catalogs to find the center, either via the dark matter (using the halo information in \texttt{halo\_NNN.hdf5}) or via the stars (using the stellar information in \texttt{star\_NNN.hdf5}).
To define the primary host, one should use the most massive halo within the zoom-in region that is uncontaminated by low-resolution dark-matter particles.
The publicly available \href{https://bitbucket.org/awetzel/halo_analysis}{\textsc{HaloAnalysis}} package for reading the halo/galaxy catalogs (see Section~\ref{sec:analysis}) automatically assigns the primary host halo(s) this way during read in (see Appendix~\ref{sec:rockstar}).
The (sub)halo catalogs also provide the best way to identify the coordinates of all other (satellite) galaxies within the zoom-in region.

Second, we more commonly use and therefore most strongly recommend an iterative zoom-in method using star particles to identify the primary galaxy.
Typically we start by measuring the mean center-of-mass position of all star particles.
We then keep those within a sphere of some large initial radius ($\approx 1 \Mpc$) around this center, and using only those star particles we recompute the center position.
We iteratively shrink this sphere by $\approx 50\%$ in radius each time and recompute the center, until the spherical radius drops below some threshold, such as $\sim 10 \pc$.
Then, we typically compute the center-of-mass velocity of all star particles within some fixed radius of this center, typically $\lesssim 8 \kpc$.
The publicly available \href{https://bitbucket.org/awetzel/gizmo_analysis}{\textsc{GizmoAnalysis}} package for reading snapshots (see Section~\ref{sec:analysis}) automatically uses this approach to assign the position and velocity of the primary galaxy(s) to the particle catalog during read in.

Furthermore, the \href{https://bitbucket.org/awetzel/gizmo_analysis}{\textsc{GizmoAnalysis}} package can identify the orientation of the galaxy, that is, the direction of the disk, if a user sets \texttt{assign\_hosts\_rotation=True} in the function \texttt{gizmo.io.Read.read\_snapshots()}.
Specifically, we first identify star particles within $10 \kpc$ of the center of the primary galaxy, and we keep only those that are within a radius that encloses 90\% of this total stellar mass, to help remove possible galaxy mergers.
Among these, we keep only the 25\% youngest, which generally are the most disk-like, and using these we measure the moment-of-inertia tensor to identify the 3 principal axes of the galaxy.
We use this moment-of-inertia tensor to rotate the coordinates into the frame of the disk.

For the \textit{Core} suite to $z\!=\!0$, we used the particle-tracking pipeline in \href{https://bitbucket.org/awetzel/gizmo_analysis}{\textsc{GizmoAnalysis}} to record the coordinates of the primary galaxy(s) at every snapshot.
Specifically, we record all star particles within the primary host halo at $z\!=\!0$, and using only these star particles that end up as part of the host today, we compute the position, velocity, and moment of inertia tensor of the primary galaxy(s) at all previous snapshots.
We store these properties in a file name \texttt{host\_coordinates.hdf5} within \texttt{track/}.
Specifically, we store the primary galaxy position and velocity in arrays named \texttt{host.position} [kpc comoving] and \texttt{host.velocity} [km s$^{-1}$], with shape $N_{\rm snapshot} \times N_{\rm host} \times N_{\rm dimension}$, where $N_{\rm snapshot}$ is the total number of snapshots (typically 600); $N_{\rm host}$ is the number of primary galaxies, which is 1 for all simulations except the \textit{ELVIS on FIRE} \ac{LG}-like simulations, for which it is 2; and $N_{\rm dimension}\!=\!3$.
The rotation tensor, named \texttt{host.rotation}, has shape $N_{\rm snapshot} \times N_{\rm host} \times N_{\rm dimension} \times N_{\rm dimension}$.

Thus, when analyzing simulations from the \textit{Core} suite, we recommend a user to read \texttt{host\_coordinates.hdf5} and use these values to locate the primary galaxy(s) at any snapshot.

\subsection{Formation coordinates and ex-situ flag for star particles}
\label{sec:exsitu}

For the \textit{Core} suite to $z\!=\!0$, the file named \texttt{host\_coordinates.hdf5} within \texttt{track/} also contains, for each star particle at $z\!=\!0$, its ``formation'' coordinates, measured at the snapshot immediately after it formed.
Given the snapshot time spacing, this is always $\lesssim 25 \Myr$ and more typically $\approx 10 \Myr$ after formation.
We measure the formation coordinates as the 3-D Cartesian x,y,z position and velocity centered on the primary galaxy and aligned with its disk orientation (principal component axes of its moment of inertia tensor, see above) at first snapshot after each star particle formed, which can be different for each star particle.
Specifically, \texttt{star.form.host.distance} [kpc physical] and \texttt{star.form.host.velocity} [km s$^{-1}$] have shape $N_{\rm particle} \times N_{\rm dimension}$, where $N_{\rm particle}$ is the number of star particles at $z\!=\!0$ and $N_{\rm dimension}\!=\!3$.
(For the \textit{ELVIS on FIRE} simulations, \texttt{host\_coordinates.hdf5} also stores \texttt{star.form.host2.distance} and \texttt{star.form.host2.velocity} for the second host galaxy.)
Thus, one can use these formation coordinate to explore how the positions and orbits of star particles at $z\!=\!0$ have changed since their formation.
One also can use them to identify star particles that formed ``ex situ", in another galaxy outside of the primary galaxy, using any desired cut on distance and/or velocity.

To make the identification of ex-situ stars even easier, for the \textit{Core} suite to $z\!=\!0$, each simulation contains a text file named \texttt{star\_exsitu\_flag\_600.txt} inside \texttt{track/} that lists, for every star particle at $z\!=\!0$, a binary flag that is 1 if the star particle formed ex situ, that is, outside of the primary galaxy in another lower-mass galaxy.
We define a star particle as ``ex-situ" following \citet{Bellardini2022}, if it formed at a spherical distance $d_{\rm form} \!>\! 30 \kpc$ comoving ($>\!30 a \kpc$ physical, where $a$ is the expansion scale factor) from the center of the primary galaxy.

\subsection{Related data sets}

The \href{https://fire.northwestern.edu}{FIRE project website}
\href{https://fire.northwestern.edu/data}{links} to several additional public data sets that relate to these FIRE-2 simulations:
\begin{itemize}

\item MUSIC cosmological initial-condition files for most of these simulations: \\
\href{http://www.tapir.caltech.edu/~phopkins/publicICs}{www.tapir.caltech.edu/$\sim$phopkins/publicICs}

\item Synthetic Gaia DR2-like surveys for 3 \ac{MW}-mass galaxies (m12i, m12f, m12m) from \citet{Sanderson2020}:
\href{http://ananke.hub.yt}{ananke.hub.yt}.\footnote{
Also at \href{http://binder.flatironinstitute.org/~rsanderson/ananke}{binder.flatironinstitute.org/$\sim$rsanderson/ananke}}
We also provide synthetic SDSS-APOGEE catalogs of radial velocities and elemental abundances \citep{Nikakhtar2021}, available as part of \href{https://www.sdss.org/dr17/}{SDSS Data Release 17} \citep{Abdurrouf2022}, query through \href{https://skyserver.sdss.org/casjobs/}{CasJobs}.\footnote{
Access at \href{http://binder.flatironinstitute.org/~rsanderson/ananke_full}{binder.flatironinstitute.org/$\sim$rsanderson/ananke\_full}}

\item Catalogs and properties of stellar streams and their progenitor galaxies for the \ac{MW}-mass simulations from \citet{Panithanpaisal2021}: \\
\href{https://flathub.flatironinstitute.org/sapfire}{flathub.flatironinstitute.org/sapfire}

\item Multipole basis expansion models of the mass distribution for the \ac{MW}-mass halos, from \citet{Arora2022}: \\
\href{https://web.sas.upenn.edu/dynamics/data/pot_models}{physics.upenn.edu/dynamics/data/pot\_models}

\item Properties of predicted binary black holes in the m12i \ac{MW}-mass galaxy from \citet{Lamberts2018}:
\href{http://ananke.hub.yt}{ananke.hub.yt}

\item Animations, images, and other visualizations: \\
\href{http://www.tapir.caltech.edu/~phopkins/Site/animations}{www.tapir.caltech.edu/$\sim$phopkins/Site/animations}

\end{itemize}

\section{Analysis Tools}
\label{sec:analysis}

The following publicly available python packages are useful for reading and analyzing these (and any) \textsc{Gizmo} simulation snapshots.
They automatically convert particle properties in a snapshot to conventional/useful units.

\begin{itemize}

\item \href{https://bitbucket.org/awetzel/gizmo_analysis}{\textsc{GizmoAnalysis}} \citep{Wetzel2020a} - use to read snapshots; analyze and visualize particle data; compute stellar evolution rates, including supernovae, stellar winds, and their nucleosynthetic yields, as used in FIRE-2; includes a Jupyter notebook tutorial: \\
\href{https://bitbucket.org/awetzel/gizmo_analysis}{bitbucket.org/awetzel/gizmo\_analysis}

\item \href{https://bitbucket.org/awetzel/halo_analysis}{\textsc{HaloAnalysis}} \citep{Wetzel2020b} - use to read and analyze halo/galaxy catalogs, generated from \textsc{Rockstar} or \textsc{AHF}, and merger trees generated from \textsc{ConsistentTrees}; includes a Jupyter notebook tutorial: \\
\href{https://bitbucket.org/awetzel/halo_analysis}{bitbucket.org/awetzel/halo\_analysis}

\item \href{https://bitbucket.org/phopkins/pfh_python}{\textsc{PFH\_python}} - use to read and analyze snapshots, including sophisticated image-making routines for generating mock Hubble-like images and movies:
\href{https://bitbucket.org/phopkins/pfh_python}{bitbucket.org/phopkins/pfh\_python}

\item \href{https://yt-project.org/}{\textsc{yt}} \citep{Turk2011, Smith2018} - a parallel-enabled simulation analysis suite with full support for reading, analyzing, and visualizing GIZMO / FIRE data, including field manipulation, particle filtering, volume rendering, and movie generation. Additional extensions including \href{http://hea-www.cfa.harvard.edu/~jzuhone/pyxsim/}{\textsc{PyXSIM}} \citep{ZuHone2016} and \href{http://trident.readthedocs.org/}{\textsc{Trident}} \citep{Hummels2017} allow for the production of synthetic observations:
\href{https://yt-project.org/}{yt-project.org}

\item \href{https://www.alexbgurvi.ch/Firefly}{\textsc{Firefly}} \citep{Geller2018} - web browser-based interactive visualization of particle-based data for science and outreach: \\
\href{https://www.alexbgurvi.ch/Firefly}{alexbgurvi.ch/Firefly}

\item \href{https://www.alexbgurvi.ch/FIRE\_studio}{FIRE Studio} \citep{Gurvich2022} - publication quality rendering of gas projection and mock Hubble stellar surface density images (including approximate dust attenuation). Also includes time-interpolation and frame centering routines for making movies in cosmological volumes. \\
\href{https://www.alexbgurvi.ch/FIRE\_studio}{alexbgurvi.ch/FIRE\_studio}

\end{itemize}

The \textsc{GIZMO} Users Guide (see Section~\ref{sec:documentation}) also lists several additional tools for analyzing and post-processing \textsc{GIZMO} snapshots, including radiative transfer, halo-finding, visualization, and other packages.

\section{Additional Documentation}
\label{sec:documentation}

\begin{itemize}

\item Video tutorial for getting started using FIRE-2 simulation data: \\
\href{https://www.youtube.com/watch?v=bl-rpzE8hrU}{www.youtube.com/watch?v=bl-rpzE8hrU}

\item FIRE project website:
\href{https://fire.northwestern.edu}{fire.northwestern.edu}

\item \textsc{GIZMO} source code (publicly available version): \\
\href{https://bitbucket.org/phopkins/gizmo-public/src/master}{bitbucket.org/phopkins/gizmo-public/src/master}

\item \textsc{GIZMO} Users Guide - provides comprehensive documentation of the \textsc{Gizmo} code and the contents of simulation snapshots:
\href{http://www.tapir.caltech.edu/~phopkins/Site/GIZMO_files/gizmo_documentation.html}{www.tapir.caltech.edu/$\sim$phopkins/Site/ \\ GIZMO\_files/gizmo\_documentation.html}

\item Meta-galactic ultraviolet background models from \citet{FaucherGiguere2009} and \citet{FaucherGiguere2020}:
\href{https://galaxies.northwestern.edu/uvb}{galaxies.northwestern.edu/uvb}

\end{itemize}

\section{License and Citing}

We release FIRE-2 data under the license \href{https://creativecommons.org/licenses/by/4.0}{Creative Commons BY 4.0}.
We request anyone using these data to cite as follows:

\textit{
We use simulations from the FIRE-2 public data release \citep{Wetzel2023}.
The FIRE-2 cosmological zoom-in simulations of galaxy formation are part of the Feedback In Realistic Environments (FIRE) project, generated using the \textsc{Gizmo} code \citep{Hopkins2015} and the FIRE-2 physics model \citep{Hopkins2018b}.
}

We also request a user to cite the individual published article(s) that introduced each simulation used, as listed in Tables 1, 2, 3, and include the URL of the FIRE project website (\href{https://fire.northwestern.edu}{fire.northwestern.edu}) in a footnote.

\section{Summary and Future Data Releases}
\label{sec:summary}

The goal of the FIRE simulation project is to develop cosmological simulations of galaxy formation that resolve the multiphase \ac{ISM} while modeling all of the major channels for stellar evolution and feedback as directly as possible, within a cosmological context.
By achieving parsec-scale resolution in cosmological zoom-in simulations, FIRE aims to improve the predictive power of galaxy formation simulations.

In this article, we described the first full public data release (DR1) of the FIRE-2 simulations,
which also represents the first public data release of a suite of cosmological zoom-in baryonic simulations across cosmic time.
This comprises 49 ``primary" galaxies in 46 different simulations across 3 suites that target different mass and redshift regimes: a \textit{Core} suite of 23 primary galaxies in 20 simulations to $z\!=\!0$, a \textit{Massive Halo} suite of 4 simulations to $z\!=\!1$, and a \textit{High Redshift} suite of 22 simulations to $z\!=\!5$.
In addition, these simulations include hundreds of resolved lower-mass (satellite) galaxies within the cosmological zoom-in regions at each snapshot.

We released full snapshots of each simulation, and we described the properties available for dark matter, stars, and gas.
We also described several additional derived data products from these simulations.
This includes accompanying (sub)halo/galaxy catalogs with member star-particle information, which allows a user to analyze not just the primary galaxy but also the many lower-mass (satellite) galaxies and dark-matter (sub)halos within each cosmological zoom-in region.
For the \textit{Core} suite, we also released, 
for each star particle at $z\!=\!0$, its formation coordinates relative to the primary galaxy(s), an ``ex-situ" flag to identify those that formed outside of the primary galaxy(s), and files of pointer indices to make it easy to track individual star particles and gas cells across snapshots.
Furthermore, for each \ac{MW}/M31-mass galaxy simulated to $z\!=\!0$, we released catalogs of stellar streams and models of the total mass distribution via multipole basis expansions.
Finally, we described how a user can access these data via FlatHUB, downloading either via a web browser or via Globus.

In Section~\ref{sec:method}, we also outlined key limitations of these FIRE-2 simulations, including physics not modeled, caveats, known tensions with observations, and subtleties of analyzing cosmological zoom-in regions.

While we released multiple snapshots for each simulation to allow users to explore redshift evolution, this DR1 includes only a subset (up to 39) of all stored snapshots (up to 600) for each simulation.
This DR1 is only the initial data release of FIRE-2 simulations, and we plan to release more data in the future, which may include more snapshots for each simulation and additional derived data products, such as full merger trees for all (sub)halos or more synthetic observations.
Future releases also may include the more recent FIRE-2 simulations that model additional physical processes, as discussed above, though these simulations encompass only a subset of all FIRE-2 galaxies.
Finally, a new suite of FIRE-3 simulations remains under active development \citep{Hopkins2023}, and we plan to release those simulations in the future as well.
We encourage users to check the FlatHUB website (\href{http://flathub.flatironinstitute.org/fire}{flathub.flatironinstitute.org/fire}) and the FIRE project website (\href{https://fire.northwestern.edu}{fire.northwestern.edu}) for the most up-to-date status of additional data releases.


\section{Acknowledgements}

We generated the FIRE-2 simulations using:
Stampede and Stampede 2, via the Extreme Science and Engineering Discovery Environment (XSEDE), supported by NSF grant ACI-1548562, including allocations TG-AST120025, TG-AST140023, TG-AST140064, TG-AST160048;
Blue Waters, supported by the NSF;
Frontera, supported by the NSF and TACC, including allocations AST21010 and AST20016;
Pleiades, via the NASA High-End Computing (HEC) Program through the NASA Advanced Supercomputing (NAS) Division at Ames Research Center, including allocations HEC SMD-16-7592, SMD-16-7561, SMD-17-120; and the Quest computing cluster at Northwestern University.
This work uses data hosted by the Flatiron Institute's FIRE data hub, and we generated data using the Flatiron Institute's computing clusters \texttt{rusty} and \texttt{popeye}; the Flatiron Institute is supported by the Simons Foundation.
\textit{yt Hub} is supported in part by the Gordon and Betty Moore Foundation's Data-Driven Discovery Initiative through Grant GBMF4561 to Matthew Turk and the National Science Foundation under Grant ACI-1535651.

AW received support from: NSF via CAREER award AST-2045928 and grant AST-2107772; NASA ATP grants 80NSSC18K1097 and 80NSSC20K0513; HST grants GO-14734, AR-15057, AR-15809, GO-15902 from STScI; a Scialog Award from the Heising-Simons Foundation; and a Hellman Fellowship.
RES and NP acknowledge support from NASA grant 19-ATP19-0068; and RES, FN, and AA acknowledge support from the Research Corporation through the Scialog Fellows program on Time Domain Astronomy, and from NSF grant AST-2007232; RES additionally acknowledges support from HST-AR-15809 from STScI.
DAA acknowledges support by NSF grants AST-2009687 and AST-2108944, CXO grant TM2-23006X, and Simons Foundation award CCA-1018464.
RF acknowledges financial support from the Swiss National Science Foundation (grant no PP00P2\_194814).
TKC is supported by the Science and Technology Facilities Council (STFC) astronomy consolidated grant ST/P000541/1 and ST/T000244/1.
JS was supported by an NSF Astronomy and Astrophysics Postdoctoral Fellowship under award AST-2102729.
ZH was supported by a Gary A. McCue postdoctoral fellowship at UC Irvine.
SL was supported by NSF grant AST-2109234 and HST-AR-16624 from STScI.
MBK acknowledges support from NSF CAREER award AST-1752913, NSF grants AST-1910346 and AST-2108962, NASA grant NNX17AG29G, and HST grants AR-15006, AR-15809, GO-15658, GO-15901, GO-15902, AR-16159, GO-16226 from STScI.
CAFG was supported by NSF through grants AST-1715216, AST-2108230, and CAREER award AST-1652522; by NASA through grants 17-ATP17-006 7 and 21-ATP21-0036; by STScI through grant HST-AR-16124.001-A; and by the Research Corporation for Science Advancement through a Cottrell Scholar Award and a Scialog Award. 
DK was supported by NSF grants AST-1715101 and AST-2108314.
Support for PFH was provided by NSF Research Grants 1911233, 20009234, 2108318, NSF CAREER grant 1455342, NASA grants 80NSSC18K0562, HST-AR-15800.

\bibliographystyle{aasjournal}
\bibliography{article}{}

\appendix

\section{\textsc{Rockstar} (sub)halo/galaxy catalogs}
\label{sec:rockstar}

Each simulation includes catalogs of (sub)halos and their galaxies at each snapshot, within a directory named \texttt{halo/}.
As Section~\ref{sec:halo} describes, we generated our default and recommended catalogs using \textsc{Rockstar-Galaxies}, using the same parameters for all simulations.
All \textsc{Rockstar} files reside in a directory \texttt{rockstar\_dm/}, named to remind the user that we ran \textsc{Rockstar} using only dark-matter particles.
By default, we store these (sub)halo catalogs in a converted HDF5 format, named \texttt{halo\_NNN.hdf5}, where \texttt{NNN} is the snapshot index, within a directory named \texttt{catalog\_hdf5/}.
We also assigned star particles to these (sub)halos in post-processing, generating corresponding galaxy stellar properties for each (sub)halo in a file named \texttt{star\_NNN.hdf5}.
(For completeness, we also provide the ASCII text files that \textsc{Rockstar} directly generates, named \texttt{out\_NNN.list}, in a directory named \texttt{catalog/}.)
Here we describe the contents of \texttt{halo\_NNN.hdf5} and \texttt{star\_NNN.hdf5}.

Again, we ran \textsc{Rockstar-Galaxies} using only dark-matter particles, so all quantities in \texttt{halo\_NNN.hdf5} are based only on dark-matter; they do not include the masses of stars or gas.
We used the halo radius definition of $\Rthm$, the radius within which the mean density of the halo is 200 times the mean matter density of the universe.

The files named \texttt{halo\_NNN.hdf5} store the following quantities for each (sub)halo:
\begin{itemize}
\item \texttt{id} - (sub)halo ID, unique only at this snapshot, indexing starts at 0
\item \texttt{id.to.index} - pointer from \texttt{id} to the array index of the (sub)halo in the catalog at this snapshot
\item \texttt{position} [kpc comoving] - 3-D position, along simulation's x,y,z coordinates
\item \texttt{velocity} [km s$^{-1}$] - 3-D velocity, along simulation's x,y,z coordinates
\item \texttt{mass} or \texttt{mass.200m} [M$_\odot$] - mass of dark matter within $\Rthm$
\item \texttt{mass.bound} [M$_\odot$] - mass of dark matter within $\Rthm$ that is bound to the (sub)halo
\item \texttt{mass.vir} [M$_\odot$] - mass of dark matter within the virial radius defined via \citet{BryanNorman1998}
\item \texttt{mass.200c} [M$_\odot$ - mass of dark matter within $\Rthc$
\item \texttt{mass.lowres} [M$_\odot$] - mass of low-resolution dark matter within $\Rthm$ as computed by \textsc{Rockstar}; this can be inaccurate in some cases, so better to use \texttt{dark2.mass} in \texttt{star\_NNN.hdf5} (see below); we recommend caution regarding any (sub)halo in which the low-resolution mass exceeds a few percent of the total mass
\item \texttt{radius} [kpc physical] - $\Rthm$
\item \texttt{scale.radius} [kpc physical] - Navarro–Frenk–White (NFW) scale radius, computed from a fit to the density profile
\item \texttt{scale.radius.klypin} [kpc physical] - NFW scale radius, computed by converting from the radius of \texttt{vel.circ.max} \citep[see][]{Klypin2011}
\item \texttt{vel.circ.max} [km s$^{-1}$] - maximum of the circular velocity profile, $\sqrt{G M_{\rm DM}(< r) / r}$
\item \texttt{vel.std} [km s$^{-1}$] - standard deviation of the velocity of member dark-matter particles
\item \texttt{axis.b/a} and \texttt{axis.c/a} - ratios of second and third to first largest ellipsoid shape axis \citep{Allgood2006}
\item \texttt{spin.peebles} - spin parameter from \citet{Peebles1969}
\item \texttt{spin.bullock} - spin parameter from \citet{Bullock2001}
\item \texttt{position.offset} [kpc physical] and \texttt{velocity.offset} [km s$^{-1}$] - offset distance and total velocity between the maximum density peak within the halo and the particle average
\end{itemize}

If using the publicly available \href{https://bitbucket.org/awetzel/halo_analysis}{\textsc{HaloAnalysis}} package to read these catalogs (which we recommend), it automatically assigns the primary host halo (which hosts the primary galaxy), defined as the most massive halo within the zoom-in region that is uncontaminated by low-resolution dark-matter particles.
\href{https://bitbucket.org/awetzel/halo_analysis}{\textsc{HaloAnalysis}} then assigns the following properties to each (sub)halo in the catalog, with respect to the center of this primary host halo.
(For the \textit{ELVIS on FIRE} \ac{LG}-like simulations, which contain two host halos, \href{https://bitbucket.org/awetzel/halo_analysis}{\textsc{HaloAnalysis}} also assigns these properties for the second host, stored as \texttt{host2.index}, \texttt{host2.distance}, and so on.)
\begin{itemize}
\item \texttt{host.index} - catalog index (not \texttt{id}) of the primary host halo at this snapshot
\item \texttt{host.distance} [kpc physical] - 3-D distance, along simulation’s x,y,z coordinates
\item \texttt{host.velocity} [km s$^{-1}$] - 3-D velocity, along simulation’s x,y,z coordinates
\item \texttt{host.velocity.tan} [km s$^{-1}$] - tangential velocity
\item \texttt{host.velocity.rad} [km s$^{-1}$] - radial velocity
\end{itemize}

For the \textit{Core} suite to $z\!=\!0$, we also generated merger trees across all 600 snapshots via \textsc{ConsistentTrees} (which we plan to release in the future), so the \texttt{halo\_NNN.hdf5} files also contain the following history-based properties (taken from the \texttt{hlist\_*.list} file that \textsc{ConsistentTrees} produces):
\begin{itemize}
\item \texttt{mass.peak} [M$_\odot$] - maximum of \texttt{mass} throughout history
\item \texttt{mass.peak.snapshot} - snapshot index when achieved \texttt{mass.peak}
\item \texttt{mass.half.snapshot} - snapshot index when first had half of \texttt{mass.peak}
\item \texttt{vel.circ.peak} [km s$^{-1}$] - maximum of \texttt{vel.circ.max} throughout history
\item \texttt{major.merger.snapshot} - snapshot index of last major merger
\item \texttt{infall.first.snapshot} - snapshot index when first became a satellite of a more massive halo
\item \texttt{infall.first.mass} [M$_\odot$] - \texttt{mass} when first became a satellite
\item \texttt{infall.first.vel.circ.max} [km s$^{-1}$] - \texttt{vel.circ.max} when first became a satellite
\item \texttt{infall.snapshot} - snapshot index when most recently became a satellite
\item \texttt{infall.mass} [M$_\odot$] - \texttt{mass} when most recently became a satellite
\item \texttt{infall.vel.circ.max} [km s$^{-1}$] - \texttt{vel.circ.max} when most recently became a satellite
\item \texttt{accrete.rate} [M$_\odot$ yr$^{-1}$] - instantaneous rate of mass accretion
\item \texttt{accrete.rate.100Myr} [M$_\odot$ yr$^{-1}$] - rate of mass growth averaged over 100 Myr
\item \texttt{accrete.rate.tdyn} [M$_\odot$ yr$^{-1}$] - rate of mass growth averaged over halo dynamical time
\end{itemize}

Finally, alongside every (sub)halo catalog file, we include a file named \texttt{star\_NNN.hdf5} that contains the following galaxy stellar properties, based on member star (or dark-matter) particles, for each halo \citep[for details on assigning member star particles, see][]{Samuel2020}:
\begin{itemize}
\item \texttt{dark2.mass} [M$_\odot$] - mass of low-resolution dark matter within $\Rthm$; this is more accurate and better to use than the \texttt{mass.lowres} that is computed by \textsc{Rockstar} in \texttt{halo\_NNN.hdf5}; we recommend caution regarding any (sub)halo in which this exceeds a few percent of the total mass
\item \texttt{star.indices} - 1-D array for each halo that lists the indices (not the \texttt{ParticleIDs}) within the simulation snapshot of member star particles
\item \texttt{star.number} - number of star particles
\item \texttt{star.mass} [M$_\odot$] - mass of all star particles
\item \texttt{star.position} [kpc comoving] - center-of-mass position of star particles, along simulation's x,y,z coordinates
\item \texttt{star.velocity} [km s$^{-1}$] - center-of-mass velocity of star particles, along simulation's x,y,z coordinates
\item \texttt{star.radius.50} [kpc physical] - radius that encloses 50\% of stellar mass
\item \texttt{star.radius.90} [kpc physical] - radius that encloses 90\% of stellar mass
\item \texttt{star.vel.std.50} [km s$^{-1}$] - velocity dispersion of star particles within \texttt{star.radius.50}
\item \texttt{star.vel.std} [km s$^{-1}$] - velocity dispersion of all star particles
\item \texttt{star.form.time.50}, \texttt{star.form.time.90}, \texttt{star.form.time.95}, \texttt{star.form.time.100} [Gyr] - age of universe when formed 50, 90, 95, 100\% of current stellar mass
\item \texttt{star.form.time.dif.68} [Gyr] - age spread between youngest 16\% and oldest 16\% of stars
\item \texttt{star.massfraction} - average across all star particles for each of the 11 \texttt{Metallicity} fields, stored as linear mass fractions as a 1-D array for each galaxy, with the following order/indices: H (0), He (1), C (2), N (3), O (4), Ne (5), Mg (6), Si (7), S (8), Ca (9), Fe (10)
\end{itemize}

\end{document}